\RequirePackage{fix-cm}
\documentclass[smallextended]{svjour3}       
\smartqed  
\usepackage{graphicx}
\usepackage{amssymb}
\usepackage{amsmath}
\usepackage{float}
\usepackage{cancel}
\usepackage{amsfonts}
\usepackage{enumitem}
\usepackage{authblk}
\usepackage[titletoc,toc,title]{appendix}
\usepackage{epstopdf}
\DeclareGraphicsExtensions{.eps}
\usepackage{braket}
\usepackage{tikz}
\usepackage{hyperref}
\hypersetup{
    colorlinks=true,
    linkcolor=blue,
    filecolor=red,      
    urlcolor=blue,
    citecolor=red
} 
\begin{document}

\title{Entanglement negativity, Holography and Black holes
}


\author{Pankaj Chaturvedi\and Vinay Malvimat \and   Gautam Sengupta 
}


\institute{Department of Physics,\\Indian Institute of Technology Kanpur,\\208016, India\at
              \email{vinaymm@iitk.ac.in}  \\         
              \email{sengupta@iitk.ac.in} \\
              \and
	  Yau Mathematical Sciences Center,\\ Tsinghua University, Beijing,\\ China\at
	      \email{cpankaj1@gmail.com}\\
}

\date{Received: date / Accepted: date}

\maketitle

\begin{abstract}

We investigate the application of our recent holographic entanglement negativity conjecture for higher dimensional conformal field theories to specific examples which serve as crucial consistency checks. 
In this context we compute the holographic entanglement negativity for bipartite pure and finite temperature mixed state configurations in $d$-dimensional conformal field theories dual to bulk pure $AdS_{d+1}$ geometry and $AdS_{d+1}$-Schwarzschild black holes respectively. It is observed that the holographic entanglement negativity characterizes the distillable entanglement for the finite temperature mixed states through the elimination of the thermal contributions.
Significantly our examples correctly reproduce universal features of the entanglement negativity for corresponding two dimensional conformal field theories, in higher dimensions.

\end{abstract}

\newpage

\section{Introduction}

The last decade has witnessed remarkable progress in the understanding of entanglement in quantum information theory and has found applications in diverse areas of theoretical physics and other related disciplines from quantum phase transitions to quantum gravity. For a bipartite $(A\cup B)$ pure state 
$|\psi_{AB}\big>$ of a quantum system  with a factorizable Hilbert space ${\cal H}={\cal H}_{A}\otimes {\cal H}_{B}$, the quantum entanglement is characterized by the {\it entanglement entropy}.
This is described by the von-Neumann entropy of the reduced density matrix $\rho_{A}=Tr_{B}\left(\rho_{A\cup B}\right)$ of the subsystem $A$ which may be computed for quantum systems with finite degrees of freedom with relative ease. On the other hand the issue of the characterization of entanglement 
for extended quantum many body systems with infinite number of degrees of freedom has proved to be extremely complex and often intractable. For 
$(1+1)$- dimensional conformal field theories ($CFT_{1+1}$) however this issue is rendered tractable through the conformal symmetry.  As demonstrated by Calabrese and Cardy 
\cite{Calabrese:2004eu,Calabrese:2009qy} in a seminal contribution, the entanglement entropy for such a $CFT_{1+1}$ may be obtained through a {\it replica technique}. This technique is based on the idea of computing the moments of the reduced density matrix $Tr(\rho^n_A)$ with $n$ being a non-negative integer or equivalently the R\'{e}nyi entropy of order $n$ which may be defined as
\begin{equation}
S^{(n)}_{A}=\frac{\ln \left[Tr\left(\rho_A^n\right)\right]}{1-n}. \label{EEodn}
\end{equation}
The quantity $Tr(\rho^n_A)$ in this computation corresponds to the partition function  on a $n$-sheeted Riemann surface with branch points at the boundaries between the subsystems $A$ and $B$ \cite{Calabrese:2004eu}. Note that the corresponding von Neumann entropy may be obtained from the above expression for the R\'{e}nyi entropy through the replica limit $n\to 1$ which has to be understood in the sense of an analytic continuation. Furthermore, the partition function for the subsystem on the $n$-sheeted  Riemann surface may be recast in terms of the correlation functions of branch-point twist fields on the complex plane \cite{Calabrese:2004eu,Calabrese:2009qy} in this limit. The corresponding correlation functions of these twist fields may then be computed directly in the $CFT_{1+1}$ to obtain the entanglement entropy.

Note that the entanglement entropy is essentially a measure for bipartite pure state entanglement. 
However, for mixed states it ceases to be a valid entanglement measure as it receives contributions from correlations irrelevant to the entanglement of the given bipartite configuration. In quantum information theory one refers to the process of {\it purification} involving a tripartition where the system being considered is embedded in a larger system in a pure state \footnote{ This procedure requires obtaining a mixed state by tracing out the degrees of freedom of a larger system in a pure state. For instance if the full system is divided in to three parts say $A_1,A_2$ and $B$ then the required density matrix $\rho_{A_1 \cup A_2}$ is obtained by tracing over the subsystem $B$.}. In a classic work Vidal and Werner \cite {PhysRevA.65.032314} introduced a computable measure termed as the {\it entanglement negativity} which characterizes the upper bound on the {\it distillable entanglement} for such a bipartite quantum system in a mixed state. This measure involves a {\it partial transpose} of the reduced density matrix over one of the subsystems in the given bipartite system. In order to define entanglement negativity it is required to consider an extended quantum system which is divided into two parts $A_1$ and $A_2$ . If $|q_i^1\big>$ and $|q_i^2\big>$ represent the bases of Hilbert space corresponding to the subsystems $A_1$ and $A_2$ respectively, then the partial transpose with respect to the degrees of freedom of the subsystem $A_2$ is expressed as 
\begin{equation}
\big<q_i^1q_j^2|\rho_{A_1 \cup A_2}^{T_{2}}|q_k^1q_l^2\big> = \big<q_i^1q_l^2|\rho_{A_1 \cup A_2}|q_k^1q_j^2\big>,\label{trace}
\end{equation}
where, $\rho_{A_1 \cup A_2}$ is the density matrix of the system $(A=A_1 \cup A_2)$. This leads to the definition of the entanglement negativity as
\begin{equation}
{\cal E} \equiv \log \big(Tr\mid\rho_{A_1 \cup A_2}^{T_{2}}\mid\big)= \log \big(Tr\mid\rho_{A}^{T_{2}}\mid\big).\label{ENeg}
\end{equation}

Observe that from the above equation, the entanglement negativity may be expressed as the logarithm over the sum of the absolute eigenvalues of the density matrix $\rho_{A}^{T_{2}}$. This may be written as follows
\begin{eqnarray}
Tr\mid\rho_{A}^{T_{2}}\mid&=&\sum_{\lambda_i >0}|\lambda_i|+\sum_{\lambda_i <0}|\lambda_i|,\label{evenTr}
\end{eqnarray}
where $\lambda_i $ correspond to the eigenvalues of the density matrix $\rho_{A}^{T_{2}}$. The entanglement negativity exhibits certain important properties including those of non-convexity and monotonicity proved by Plenio in \cite{Plenio:2005cwa}. 

Recently, the issue of obtaining the entanglement negativity in $(1+1)$-dimensional conformal field theories has received considerable attention. In \cite{Calabrese:2012ew,Calabrese:2012nk,Calabrese:2014yza} the authors have advanced a systematic procedure for this which involves the
{\it replica technique} mentioned earlier, to compute the entanglement negativity by relating it to the appropriate correlation functions of the twist fields. Through this procedure, the authors were able to demonstrate that the {\it entanglement negativity} precisely characterizes the
upper bound on the distillable entanglement.

In \cite {Ryu:2006bv,Ryu:2006ef} Ryu and Takayanagi conjectured a holographic prescription in the context of the $AdS/CFT$ correspondence which leads to the
entanglement entropy in $d$-dimensional holographic conformal field theories.
Their prescription for the entanglement entropy $S_A$ of a spatial region $A$ (enclosed by the boundary $\partial A$) involves the area of the  minimal surface (denoted by $\gamma_A$) extending into the $(d+1)$-dimensional bulk and anchored on the subsystem $A$ as follows
\begin{equation}
S_A=\frac{Area(\gamma_A)}{(4G^{(d+1)}_N)},\label{EEI}
\end{equation}
where, $G_N^{(d+1)}$ is the gravitational constant of the bulk space time. Application of this holographic prescription to compute the entanglement entropy for various holographic CFTs  has yielded interesting insights  \cite{Takayanagi:2012kg,Calabrese:2009nk,Cadoni:2009tk,e12112244,Hubeny:2012ry,Nishioka:2009un,Fischler:2012ca,Fischler:2012uv,Chaturvedi:2016kbk} ( and references therein).

From the above discussion it is evident that a holographic description in the context of the $\it{AdS}/\it{CFT}$ correspondence, for the entanglement negativity in conformal field theories is a critical open issue. In this context, in \cite{Rangamani:2014ywa} the authors have computed the entanglement negativity for the pure state described by the vacuum of a conformal field theory which is  dual to the bulk pure $\it{AdS}$ spacetime. Furthermore in \cite{Banerjee:2015coc} the authors have conjectured a generalized holographic $c$-function which in the dual CFT may correspond to some mixed state entanglement measure.

Very recently we have proposed a holographic entanglement negativity conjecture for bipartite pure and mixed states of a
holographic CFT \cite{Chaturvedi:2016rcn} in the $AdS_3/CFT_2$ scenario. Interestingly, the holographic entanglement negativity may be described through an algebraic sum of the lengths of space like geodesics
anchored on appropriate intervals in the dual CFT. Curiously this reduces to a specific sum of the {\it holographic mutual informations} between the intervals in question, upto a numerical factor \footnote{Note that entanglement negativity and mutual information are completely distinct measures in quantum information theory. However their universal parts which are dominant in the holographic (large central charge) limit  match for the bipartite configuration in question. See the end of section \ref{3} for a more detailed discussion regarding this issue.}. Our holographic conjecture exactly reproduces the 
the universal part of the corresponding replica technique results for the dual CFT described in \cite{Calabrese:2014yza},
in the large central charge limit for the following bipartite pure and mixed state configurations. These involve
the pure vacuum state and the finite temperature mixed state configurations dual to bulk pure $AdS_3$ space-time and the bulk Euclidean BTZ black hole respectively. The results for the configurations mentioned above are
strongly substantiated by a large central charge analysis for the entanglement negativity of a holographic $CFT_{1+1}$, utilizing the monodromy technique as described in \cite{Malvimat:2017yaj}. We mention here that despite these significant consistency checks, a bulk proof for our conjecture along the lines of \cite {Faulkner:2013yia} remains a critical open issue to be addressed.

Our holographic entanglement negativity conjecture  for bipartite quantum states of a $CFT_{1+1}$ in the $AdS_3/CFT_2$ scenario naturally suggests a higher dimensional extension following \cite{Ryu:2006bv,Ryu:2006ef} in a more generic 
$AdS_{d+1}/CFT_d$ scenario, alluded to in \cite{Chaturvedi:2016rcn}. As described there the higher dimensional
extension involves an algebraic sum of the areas of bulk static minimal surfaces anchored on appropriate boundary subsystems which is again proportional to a specific sum of the {\it holographic mutual information} between appropriate subsystems. Note that the higher dimensional extension of our conjecture necessitates a formal bulk proof along the lines of \cite{Lewkowycz:2013nqa}, which remains a non trivial open issue. Hence it is important to first establish consistency checks through the application of the conjecture to specific higher dimensional examples in order to investigate the reproducibility of universal features of entanglement negativity for $CFT_{1+1}$ described in \cite {Calabrese:2014yza,Chaturvedi:2016rcn}. Such an exercise is expected to provide crucial insights into the higher dimensional extension and to a possible proof for the conjecture.
   
In this article we address the above issue and apply our holographic conjecture in \cite{Chaturvedi:2016rcn} (CMS) to compute the entanglement negativity for bipartite pure and mixed states of specific higher dimensional CFTs. These involve the pure vacuum state of a $CFT_d$ dual to a bulk pure $AdS_{d+1}$ space-time and finite temperature mixed state dual to a $AdS_{d+1}$-Schwarzschild black hole. These examples lead to extremely interesting
results described below. We observe that for the pure state described by the $CFT_d$ vacuum, the holographic entanglement negativity is proportional to the holographic entanglement entropy. It is further observed
that the holographic entanglement negativity characterizes the upper bound on the distillable entanglement for the finite temperature mixed state of the $CFT_d$, through the elimination of the thermal contributions. Remarkably the above results following from our conjecture, constitute the exact reproduction of the universal features of entanglement negativity in $CFT_{1+1}$ described in \cite{Calabrese:2012ew,Calabrese:2012nk,Calabrese:2014yza}, for higher dimensional holographic $CFT_d$. Quite evidently the above results constitute strong consistency checks for the higher dimensional extension of our conjecture despite the absence of a formal bulk proof.

This article is organized as follows. In Section 2, we briefly collect the results in \cite {Calabrese:2014yza} for the entanglement negativity of both pure and mixed states in a $CFT_{1+ 1}$ which is reviewed in the Appendix. Subsequently in Section 3, we briefly describe our conjecture in the context of the $AdS_3/CFT_2$ scenario \cite {Chaturvedi:2016rcn} (CMS) and its subsequent generalization to the $AdS_{d+1}/CFT_d$ framework. In Section 4, we employ our holographic  conjecture to obtain the entanglement negativity for both pure and mixed states in holographic $CFT_d$ involving a subsystem with rectangular strip geometry. In the last section we provide a summary of our results and discuss future open issues.

\section{Entanglement entropy and entanglement negativity in $CFT_{1+1}$}

In this section we begin by briefly reviewing the procedure for computing the entanglement entropy for bipartite pure and finite temperature mixed states of a $CFT_{1+1}$ and discuss its inadequacy as an entanglement measure for the mixed states. Subsequently we briefly outline the results for the entanglement negativity of both pure and finite temperature mixed states in a $CFT_{1+1}$. This is reviewed in detail in the Appendix.

\subsection{Entanglement entropy}
For an extended bipartite quantum system which is bipartitioned into a subsystem $A$ and it's complement $A^c$, the entanglement entropy corresponding to the subsystem $A$ is given as
\begin{equation}
S_{A}=\lim_{n \to 1}\frac{\ln (Tr[\rho_A^n])}{1-n}=-\lim_{n \to 1} \frac{\partial}{\partial n}Tr[\rho_A^n],\label{EECFT1}
\end{equation}
where, $\rho$ is the full density matrix and $\rho_A=Tr_{A^{c}}(\rho)$ denotes the reduced density matrix for
the subsystem-$A$ and $n\to 1$ is the replica limit. For a $CFT_{1+1}$, the moments of the reduced density matrix $Tr(\rho_{A}^n)$ are related to the partition function on  a $n$-sheeted Riemann surface with branch points at the boundaries between regions $A$ and $A^c$ \cite{Calabrese:2004eu}. Alternatively, the partition function on a $n$-sheeted Riemann surface may be recast as the correlation function of the branch-point twist/anti-twist fields ${\cal T}_{n}$ and $\overline{{\cal T}}_n$ on the complex plane with the following scaling dimensions 
\begin{equation}
\Delta_n=\frac{c}{12}(n-1/n),\label{dimTn}
\end{equation}
here, $c$ is the central charge of the CFT. Hence following \cite{Calabrese:2004eu,Calabrese:2009qy} the general form for the quantity $Tr\rho_A^n$ may be expressed as follows
\begin{equation}
Tr\rho_A^n=\big<{\cal T}_n(u_1)\overline{{\cal T}}_n(v_1)\cdots{\cal T}_n(u_N)\overline{{\cal T}}_n(v_N)\big>,\label{rhoAn}
\end{equation} 
where, $A=\cup_{i=1}^N[u_i,v_i]$ indicates that the subsystem $A$ has been divided into 
$N$ disjoint intervals. For the case when $N=1$ with the subsystem length $|u-v|=\ell$, the eq.(\ref{rhoAn}) reduces to the following 
\begin{equation}
Tr\rho_A^n=\big<{\cal T}_n(u)\overline{{\cal T}}_n(v)\big>=c_n\left(\frac{\ell}{a}\right)^{-c/6(n-1/n)},\label{rhoAn1}
\end{equation}
here, $c_n$ is some constant and $a$ is the $UV$ cut-off  for the $(1+1)$-dimensional CFT. The expression for the entanglement entropy in eq.(\ref{EECFT1}) along with the eq.(\ref{rhoAn1}) leads to the following result
\begin{equation}
S_A=\frac{c}{3}\ln\left(\frac{\ell}{a}\right)+constant.\label{2DEE}
\end{equation}
The above result corresponds to the entanglement entropy of a subsystem $A$ with length 
$\ell$ for the $CFT_{1+1}$ vacuum. The corresponding result for the finite temperature mixed state requires the evaluation 
of the two point twist correlator in eq. (\ref {rhoAn1}) on a cylinder of circumference $\beta= 1/T$\cite{Calabrese:2004eu,Calabrese:2009qy}. The above procedure leads to the following expression for the entanglement entropy of the subsystem $A$ as
\begin{equation}
S_A= \frac{c}{3}\log \left(\frac{\beta}{\pi a} \sinh \frac{\pi \ell}{\beta} \right)+constant.\label{2DEET}
\end{equation}
 Observe that from eq.(\ref {2DEET}) the large temperature limit leads to the purely thermal entropy indicating that the entanglement entropy receives contribution from both the classical ( thermal) and the quantum correlations at finite temperatures. A similar observation may also be made for the case of finite temperature mixed states of higher dimensional conformal field theories which are  dual to bulk $AdS$ black holes in the context of the Ryu and Takayanagi conjecture \cite{Fischler:2012ca,Chaturvedi:2016kbk}. This is a generic issue in quantum information theory and hence the entanglement entropy
ceases to be valid measure to characterize mixed state entanglement. This naturally leads to the question of establishing appropriate measures to characterize the distillable quantum entanglement for a mixed state which in this case is described by a finite temperature $CFT$. As mentioned earlier this issue may be addressed through the entanglement negativity measure introduced by Vidal and Werner \cite {PhysRevA.65.032314}. We now proceed to describe the computation of the entanglement negativity for bipartite pure and mixed states of a $CFT_{(1+1)}$.

\subsection{Entanglement negativity in $CFT_{(1+1)}$}
In order to define entanglement negativity in $(1+1)$-dimensional CFTs it is required to consider 
the tripartition $A_1$,$A_2$ and $A^c$ such that $A_1$ and $A_2$ correspond to finite intervals $[u_1,v_1]$ and 
$[u_2,v_2]$ of lengths $\ell_1$ and $\ell_2$ respectively whereas, $A^c$ represents the rest of the system. 
Let $\rho_A$ denote the reduced density matrix of the subsystem $A=A_1\cup A_2$ such that $\rho_A=\rho_{A_{1} \cup A_2}$ which is obtained by tracing out the full density matrix $\rho$ over the part $A^c$, i.e. 
$\rho_A=Tr_{A^c}\left(\rho\right)$. As mentioned earlier in the Introduction, the entanglement negativity is then given by
 eq.(\ref {ENeg}). The authors in \cite{Calabrese:2014yza} employed the replica technique to show that the entanglement negativity ${\cal E}$ for $(1+1)$-dimensional CFTs may be expressed as follows
\begin{equation}
{\cal E}=\lim_{n_e \to 1}\ln\left[Tr(\rho^{T_A})^{n_e}\right].\label{ENCFTI}
\end{equation}
Note that in the above equation $\rho=\rho_{A\cup A^c}$ corresponds to the full density matrix. The replica limit $n_e\to 1$ indicates that negativity is defined as an analytic continuation\footnote{\label{note1}Note that a complete general construction for this analytic continuation is still an open issue  although this has been addressed for some explicit examples of simple conformal field theories in \cite{Calabrese:2009ez,Calabrese:2010he,Calabrese:2014yza} ( see also Headrick \cite{Headrick:2010zt} ) and also in the condensed matter physics literature.} of an even sequence of  $n$ ($n_e$ represents even values of n) to $n_e= 1$.  The computational details of the transition from a tripartite configuration $(A_1,A_2,A^c)$ to a bipartite configuration $(A,A^c,\cancel{0})$ are reviewed in the Appendix.

It follows that the entanglement negativity for the bipartite pure state described by the $CFT_{1+1}$ vacuum is obtained  through a specific two point twist correlator as follows
\begin{equation}
 {\cal E}=\lim_{n_e \to 1}\ln\bigg[
 \langle{\cal T}^2_{n_e}(u)\overline{{\cal T}}^2_{n_e}(v)\rangle_{\mathbb{C}}\bigg]\label{ENCFTP}
\end{equation}
As demonstrated by authors in \cite{Calabrese:2012ew,Calabrese:2012nk}, the twist fields ${{\cal T}}^2_{n_e}$ connect $n_e^{th}$ sheet of the Riemann surface to $(n_e+2)^{th}$ sheet of the Riemann surface whereas the twist field $\overline{{\cal T}}^2_{n_e}$ connects $n_e^{th}$ sheet to $(n_e-2)^{th}$ sheet of the Riemann surface. This led the authors to conclude that the  the correlator in eq.(\ref{ENCFTP}) factorizes due to the decoupling of $n_e$ even sheeted Riemann surface into two $n_e/2$ sheeted Riemann surfaces as follows
\begin{equation}
 \langle{\cal T}^2_{n_e}(u)\overline{{\cal T}}^2_{n_e}(v)\rangle_{\mathbb{C}}=\langle{\cal T}_{\frac{n_e}{2}}(u)\overline{{\cal T}}_{\frac{n_e}{2}}(v)\rangle_{\mathbb{C}}^2.\label{42f}
\end{equation} 
Therefore, the scaling dimension 
$(\Delta_{n_e}^{(2)})$ of the operator ${\cal T}_{n_e}^2$ may be related to the scaling dimensions 
$(\Delta_{n_e})$ of the operator ${\cal T}_{n_e}$  as follows
\begin{eqnarray}
\Delta_{n_e}^{(2)}=2 \Delta_{n_e/2}&=&\frac{c}{6}\left(\frac{n_e}{2}-\frac{2}{n_e}\right),\label{sctn2}\\
\Delta_{n_e}&=& \frac{c}{12}\left(n_e-\frac{1}{n_e}\right).\label{dimTn2}
\end{eqnarray}
Utilizing the well known form for the two point twist correlator given in eq.(\ref{42f}) and substituting it in eq.(\ref{ENCFTP}), one arrives at the following result
\begin{equation}
{\cal E}=\frac{c}{2}\ln\left(\frac{\ell}{a}\right)+constant=\frac{3}{2}S_A + const. \label{ENCFTzeroI}
\end{equation}
The result matches with the expectation from quantum information theory that the entanglement negativity for a pure state is the R\'enyi-entropy of order half and for the pure vacuum state of the $CFT_{1+1}$ the universal part is  proportional to the entanglement entropy. Furthermore, the authors also showed that for the finite temperature
mixed state, the entanglement negativity is related to a specific four point twist correlator as follows
\footnote{Note that for the finite temperature mixed state the quantity $Tr(\rho^{T_A})^{n_e}$ is not described by the twist correlator in eq.(\ref{42f}) because of certain geometrical reasons described in \cite{Calabrese:2014yza} (See also section \ref{bfms} of the Appendix).}
\begin{equation}
 {\cal E}=\lim_{L \to \infty}\lim_{n_e \to 1}\ln\left[\big<{\cal T}_{n_e}(-L)\overline{{\cal T}}^2_{n_e}(-\ell){\cal T}^2_{n_e}(0)\overline{{\cal T}}_{n_e}(L)\big>_{\beta}\right],\label{ENCFTM}
\end{equation}
where the subscript $\beta$ indicates that the above four point function has to be computed for a finite temperature on an infinite cylinder with circumference $\beta$.
Evaluating the four point function given in eq.(\ref{ENCFTM}) it could be shown that the entanglement negativity for the finite temperature mixed state may be expressed as
\begin{equation}
 {\cal E}=\frac{c}{2}\bigg[\ln\bigg\{\frac{\beta}{\pi a}\sinh(\frac{\pi\ell}{\beta})\bigg\}-\frac{\pi \ell}{\beta}\bigg]+f(e^{-2\pi \ell/\beta})+\ln (c^2_{1/2} c_1).\label{ENCFTfiniteI}
\end{equation}
Here $c_{1/2}$ and $c_1$ are the normalization  constants for the two-point twist correlators (See Appendix for details of the above computations). The function $f(x)$ where $x=e^{-2\pi \ell/\beta}$ and the constants are non universal and depend on the full operator content of the theory. For brevity the above 
eq.(\ref {ENCFTfiniteI}) may be re-expressed as
\begin{equation}
{\cal E}=\frac{3}{2}\bigg[S_A-S_A^{th}\bigg]+f(e^{-2\pi \ell/\beta})+const,\label{ENCFTfiniteu}
\end{equation}
where $S_A=\frac{c}{3}\ln\left[\frac{\beta}{\pi a}\sinh\left(\frac{\pi \ell}{\beta}\right)\right]$ 
corresponds to the entanglement entropy and $S_A^{th}=\frac{\pi c \ell}{3\beta}$ to the thermal entropy of the subsystem-$A$. This is an extremely significant result illustrating that  for the finite temperature mixed state of a $CFT_{1+1}$,  the negativity ${\cal E}$  characterizes the upper bound on the distillable entanglement through the elimination of the thermal contributions. In the next subsection, we discuss the large central charge $(c)$ limit of the above result and its significance in the context of the $AdS/CFT$ correspondence.

\subsection{Large central charge limit of the entanglement negativity in $CFT_{1+1}$}

In this section, we discuss the the large central charge limit ($c\rightarrow \infty$) of the four point twist correlator 
which is related to the entanglement negativity for the bipartite finite temperature mixed state of a $CFT_{1+1}$, mentioned in eq.(\ref{ENCFTfiniteu}). To this end consider a four point function of the primary operators ${\cal O}_i$ inserted at  points $z_i\ (i=1,2,3,4)$ on the complex plane, and their corresponding scaling dimensions denoted by $\Delta_i$. Under the conformal transformation $w=\frac{(z-z_1)(z_3-z_4)}{(z-z_4)(z_3-z_1)}$, the four point function may be expanded in terms of the conformal blocks as follows
\begin{equation}
  \big<{\cal O}_1(0){\cal O}_2(x) {\cal O}_3(1){\cal O}_4(\infty)\big> =\sum_{p}a_p \varPsi(h_i,h_p,x) \overline{\varPsi}(\overline{h}_i, \overline{h}_p,\overline{x}).\label{gx1}
\end{equation}
Here  $x$ is the cross ratio given by $x=\frac{z_{12}z_{34}}{z_{13}z_{24}}$, $h_i$ and $\bar{h}_i$ are the holomorphic and the anti-holomorphic scaling dimensions of the operation ${\cal O}_i$. The summation in the above equation is over all the primary operators 
$ {\cal O}_p$ with scaling dimensions $h_p$ and $\bar{h}_p$. $\varPsi(h_i,h_p,x)$ and $\overline{\varPsi}(\overline{h}_i, \overline{h}_p,\overline{x})$ are the corresponding conformal blocks. 

In recent years, there has been significant effort to determine the large central charge limit of the above mentioned conformal blocks. Although there is no rigorous proof for this, there is strong evidence that these blocks exponentiate in the limit $c\to\infty$ (as long as $\frac{h_i}{c}$ and $\frac{h_p}{c}$ are held fixed in this limit) \cite{Fitzpatrick:2014vua,Hartman:2013mia}. This exponentiation may be expressed as follows
\begin{equation}
 \varPsi(h_i,h_p,x)\approx \exp{\bigg[-\frac{c}{6}~g(\frac{h_i}{c},\frac{h_p}{c} ,x)\bigg]}.\label{4ptl}
\end{equation}
Note that this result is valid in the large central charge limit alone and there are both perturbative and non-perturbative corrections in $O[\frac{1}{c}]$. The method to determine the exponentiated blocks involves examining their monodromy properties around the singularities of the stress tensor $T(z)$ in various channels. This technique is based on earlier works by Zamolodchikov et al. where they had examined the semi-classical conformal blocks in the context of the Liouville field theory \cite{Zamolodchikov:1985ie,Zamolodchikov:1995aa,Harlow:2011ny}.

The above mentioned technique has been used to investigate the large central charge limit of the entanglement entropy of two disjoint intervals in a $CFT_{1+1}$ which is also described by a specific four point twist correlator \cite{Headrick:2010zt,Hartman:2013mia,Faulkner:2013yia,Banerjee:2016qca}. In these articles the authors have shown that that the leading large central charge contribution to the corresponding four point function is universal (i.e it is independent of the operator content of the theory) and matches exactly with that predicted from the Ryu-Takayanagi conjecture.

Observe that the above arguments also apply to to the four point function twist correlator in a $CFT_{1+1}$ that is related to the entanglement negativity in eq. (\ref {ENCFTM}) \footnote{\label{foot5} In a recent article (arXiv:1712.02288) utilizing the monodromy technique, we have provided a proof of this assertion for the four point function related to the entanglement negativity. Also note that for a simpler case of a mixed state described by two adjacent intervals in a $CFT_{1+1}$  the large central charge result for the entanglement negativity was obtained in \cite{Kulaxizi:2014nma} which bears out the above assertion.}. Hence we expect that in the large central charge limit the non-universal term given by the function $f(x)$ in eq.(\ref{ENCFTfiniteI}) for the entanglement negativity is sub leading and the leading contribution arises from the universal part which is expressed below 
\begin{equation}
{\cal E}=\frac{c}{2}\bigg[\ln\bigg\{\frac{\beta}{\pi a}\sinh(\frac{\pi\ell}{\beta})\bigg\}-\frac{\pi \ell}{\beta}\bigg].\label{larcEN} 
\end{equation}
From the above discussion it is clear that in the large central charge limit, the entanglement negativity for the bipartite finite temperature  mixed state of a $CFT_{1+1}$ assumes this universal form illustrating the elimination of the thermal contribution and leading to the distillable entanglement. In the context of the $AdS/CFT$ correspondence, the large central charge limit essentially describes the large $N$ limit of the boundary CFT through the Brown-Henneaux formula \cite {Henningson:1998gx,Karch:2015rpa}.  This leads us to the possibility of a corresponding holographic conjecture for the entanglement negativity in the $AdS/CFT$ scenario. As mentioned earlier, in \cite{Chaturvedi:2016rcn} (CMS) we proposed such a holographic conjecture which exactly reproduces the above result in eq. (\ref {larcEN}) from a bulk computation which involves a Euclidean BTZ black hole in the $AdS_3/CFT_2$ scenario. Furthermore we also demonstrated that our conjecture leads to the correct form for the negativity of a bipartite pure state 
described by the $CFT_{1+1}$ vacuum given in eq.(\ref{ENCFTzero}). This is briefly reviewed in the following section.

\section{ Holographic prescription for the entanglement negativity}\label{3}

\begin{figure}[H]
\centering
\includegraphics[width =3in,height=2.2in]{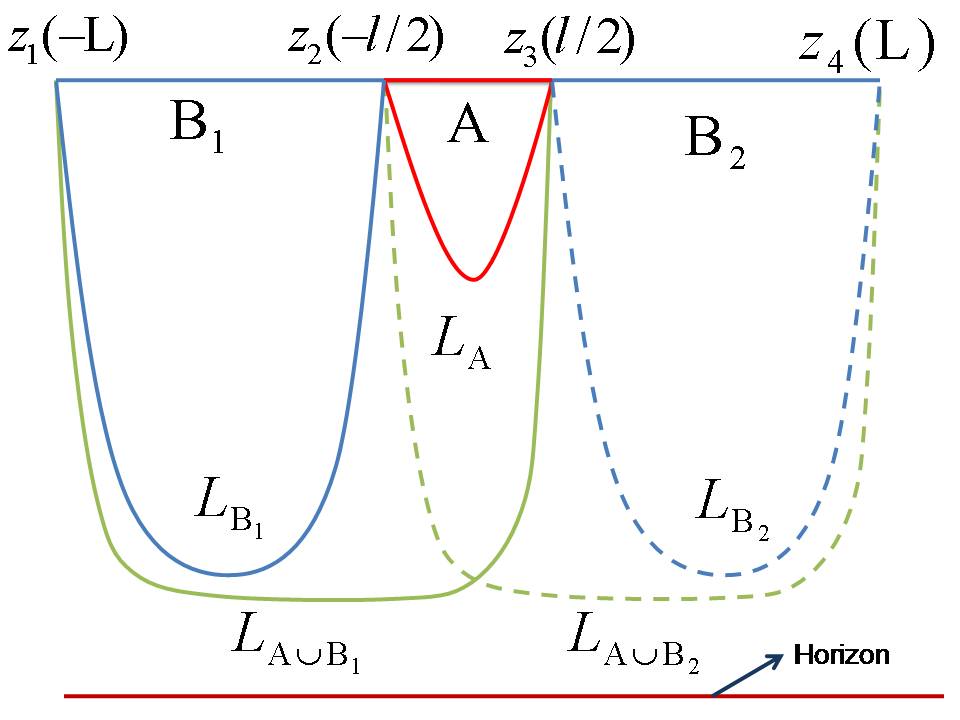}
\caption{\label{fig:subsystem} Schematic of geodesics anchored on the subsystems $A$, $B_1$ and $B_2$ in the dual $CFT_{1+1}$, which are relevant for our holographic conjecture.}\end{figure}

In this section, we review the holographic prescription proposed in \cite{Chaturvedi:2016rcn} (CMS) for the entanglement negativity of a bipartite ($A\cup A^c$) quantum states of a $CFT_{1+1}$ in the $AdS_3/CFT_2$ scenario. To begin with let us consider the dual $CFT_{1+1}$ to be partitioned into the subsystem $A$ and its complement $A^c$. We denote $B_1$ and $B_2$ as two large finite intervals adjacent to $A$ on either side of it such that $B=B_1\cup B_2$ as shown in fig.(\ref{fig:subsystem}). As mentioned in section-2, the entanglement negativity is defined in the limit $B\to A^c$ ( $L\to\infty$) which corresponds to extending the subsystems $B_1$ and $B_2$ to infinity.

The form of the two point twist correlators in a $CFT_{1+1}$ may be expressed as follows
\begin{eqnarray}
\big<{\cal T}_{n_e}(z_k)\overline{{\cal T}}_{n_e}(z_l)\big>_{\mathbb{C}}&=& \frac{c_{n_e}}{z_{kl}^{2\Delta_{n_e}}}~~~~~~~~~~~~~~~\label{2pc2}\\
\langle{\cal T}^2_{n_e}(u)\overline{{\cal T}}^2_{n_e}(v)\rangle_{\mathbb{C}}&=&\big<{\cal T}_{\frac{n_e}{2}}(z_i)\overline{{\cal T}}_{\frac{n_e}{2}}(z_j)\big>^2_{\mathbb{C}}= \frac{c_{n_e/2}^2}{z_{ij}^{4\Delta_{\frac{n_e}{2}}}},~~~~\label{2pc1}
\end{eqnarray} 
where, we have used the factorization given in eq.(\ref{42f}), $z_{ij}=|z_i-z_j|$ and $c_{n_e}$ is the normalization constant. Observe that the universal part of the required four point twist correlator\footnote{Note that for the mixed state depicted in fig.(\ref{fig:subsystem}) the four point function  has to be evaluated on the complex plane for the  $CFT_{1+1}$ vacuum whereas for the finite temperature case it has to be evaluated on an infinite cylinder. See eq.(\ref{t4pt}) in appendix-\ref{FTEN} for the transformation relating the two.}  given by eq.(\ref{rhoAn5}) in the appendix \ref{FTEN}, that provides the dominant contribution in the large central charge factorizes as follows
\begin{align}
 \big< {\cal T}_{n_e}(z_1)\overline{{\cal T}}^2_{n_e}(z_2){\cal T}^2_{n_e}(z_3)\overline{{\cal T}}_{n_e}(z_4)\big>_{\mathbb{C}}&=
 \big<{\cal T}_{\frac{n_e}{2}}(z_2)\overline{{\cal T}}_{\frac{n_e}{2}}(z_3)\big>^2\big<{\cal T}_{n_e}(z_1)\overline{{\cal T}}_{n_e}(z_4)\big>\nonumber\\&\times \frac{\big<{\cal T}_{\frac{n_e}{2}}(z_1)\overline{{\cal T}}_{\frac{n_e}{2}}(z_2)\big>\big<{\cal T}_{\frac{n_e}{2}}(z_3)\overline{{\cal T}}_{\frac{n_e}{2}}(z_4)\big>}{\big<{\cal T}_{\frac{n_e}{2}}(z_1)\overline{{\cal T}}_{\frac{n_e}{2}}(z_3)\big>\big<{\cal T}_{\frac{n_e}{2}}(z_2)\overline{{\cal T}}_{\frac{n_e}{2}}(z_4)\big>}\nonumber\\&~~~ +O[\frac{1}{c}]\label{factor}
\end{align} 
Note that as discussed in the previous section the sub leading non-universal term that depends on the full operator content of the theory, given by the function  $f(x)=\underset{n_e\to1}{\lim}\ln[{\cal F}_{n_e}(x)]$ has been neglected  in the semi classical large central charge limit $(c\to \infty)$ in the above equation.
From the $AdS/CFT$ dictionary the two point functions in eq.(\ref{2pc1}) and eq.(\ref{2pc2}) on the boundary $CFT_{1+1}$ may be related to the length of the geodesic ${\cal L}_{ij}$ anchored on the points $(z_i,z_j)$ and extending into the bulk $AdS_{2+1}$  space time as follows
\begin{eqnarray}
\big<{\cal T}_{n_e}(z_k)\overline{{\cal T}}_{n_e}(z_l)\big>_{\mathbb{C}} &\sim& e^{-\frac{\Delta_{n_e}{ \cal L}_{kl}}{R}}\label{l34}\\
\big<{\cal T}_{\frac{n_e}{2}}(z_i)\overline{{\cal T}}_{\frac{n_e}{2}}(z_j)\big>_{\mathbb{C}} &\sim& e^{-\frac{\Delta_{\frac{n_e}{2}}{ \cal L}_{ij}}{R}},\label{l12}
\end{eqnarray}
where, $R$ is the $\it{AdS}$ radius of the bulk $AdS_{2+1}$ space time. From fig.(\ref{fig:subsystem}) one may identify that 
\begin{eqnarray}
&{\cal L}_{12}={\cal L}_{B_1},~~{\cal L}_{23}={\cal L}_{A},~~{\cal L}_{34}={\cal L}_{B_2},\nonumber \\&~~{\cal L}_{13}={\cal L}_{A\cup B_1},~~ {\cal L}_{24}={\cal L}_{A\cup B_2},~~{\cal L}_{14}={ \cal L}_{A\cup B}.~~\label{Lrel}
\end{eqnarray}

With the identification in eq.(\ref{Lrel})and substituting eq.(\ref{l12}) and eq.(\ref{l34}) in , reduces to the following form in terms of the geodesic lengths as 
\begin{eqnarray}
\big<&{\cal T}_{n_e}(z_1)\overline{{\cal T}}^2_{n_e}(z_2){\cal T}^2_{n_e}(z_3)\overline{{\cal T}}_{n_e}(z_4)\big>_{\mathbb{C}}\sim \exp{[\frac{-\Delta_{n_e} X-\Delta_{\frac{n_e}{2} } Y}{R}]},\label{rhoAn7}
\end{eqnarray}
where
\begin{eqnarray}
X&=&{ \cal L}_{A\cup B} \\Y&= &2{\cal L}_{A}+{\cal L}_{B_1}+{\cal L}_{B_2}-{\cal L}_{A\cup B_1}-{\cal L}_{A\cup B_2}
\end{eqnarray}

 From eq.(\ref{sctn2}) and eq.(\ref{dimTn2}),observe that in the replica limit\footnote{Note that the negative scaling dimension in the replica limit has to be understood only in the sense of analytic continuation. Construction of such an analytic continuation is an extremely complex problem. See also footnote(\ref{note1}).} $n_e\rightarrow 1$, we have $\Delta_{n_e}\rightarrow 0$ and $\Delta_{n_e}^{(2)}\rightarrow -\frac{c}{4}$ . It is also to be noted that the central charge `$c$' of $CFT_{1+1}$ is related to the AdS length $R$ through the Brown-Henneaux formula $c=\frac{3R}{2G_{N}^{3}}$, where $G_{N}^{3}$ is the $(2+1)$-dimensional gravitational  constant\cite{brown1986}. Therefore, utilizing the above mentioned Brown-Henneaux formula, eq.(\ref{rhoAn7}) and eq.(\ref{factor}) one may express the holographic entanglement negativity for the bipartite system ($A\cup A^c$) as follows
\begin{equation}
{\cal E}= \lim_{B \to A^c}\frac{3}{16G_{N}^{3}}\bigg[(2 {\cal L}_{A}+{\cal L}_{B_1}+{\cal L}_{B_2}-{\cal L}_{A\cup B_1}-{ \cal L}_{A\cup B_2})\bigg].\label{ENCFT3}
\end{equation}
 In the $AdS_3/CFT_2$ scenario the Ryu and Takayanagi conjecture relates the geodesic length to the entanglement entropy as given in eq.(\ref{EEI}). This enables us to express the above eq.(\ref{ENCFT3}) which describes our holographic conjecture for the entanglement negativity as follows
\begin{equation}
{\cal E}=\lim_{B \to A^c}\frac{3}{4}\bigg[2 S_{A}+S_{B_1}+S_{B_2}-S_{A\cup B_1}-S_{A\cup B_2}\bigg],\label{ENCFTSA}
\end{equation}
Note that the holographic mutual information between the pair of intervals $(A,B_i)$($i=1,2$)  as follows
\begin{eqnarray}
 {\cal I}(A,B_i)&=& S_{A}+S_{B_i}-S_{A\cup B_i},\nonumber\\
&=&\frac{1}{4G_{N}^{(3)}}({\cal L}_{A}+{ \cal L}_{B_i}-{\cal L}_{A\cup B_i}),~~~~~~\label{MI}
\end{eqnarray}
Quite interestingly, using eq.(\ref{MI}) in eq.(\ref{ENCFTSA}) we may re-express our conjecture in terms of the holographic mutual information as
\begin{eqnarray}
{\cal E}&=&\lim_{B\to A^{c}} \frac{3}{4}\big[{\cal I}(A,B_1)+{\cal I}(A,B_2)\big],\label{ENCFT4}
\end{eqnarray}
It is to be emphasized  here that the mutual information and the entanglement negativity are distinct quantum information theoretic measures. Entanglement negativity is the upper bound on the distillable entanglement whereas the mutual information is the upper bound on the total correlations of a bipartite system. However in the large central charge limit their leading universal parts match exactly for the bipartite configuration in question whereas the sub leading non universal terms are distinct. This matching between the universal parts of these two measures has also been observed for both global and local quench for the case of the mixed state of adjacent intervals in a $CFT_{1+1}$ \cite{Coser:2014gsa,Wen:2015qwa}. Choosing the corresponding subsystems as shown in the fig.(\ref{fig:subsystem}), the eq.(\ref{ENCFT3}) may now be used to compute the entanglement negativity of the bipartite systems described by $(1+1)$-dimensional boundary CFT purely in terms of the bulk quantities. In the next section we will briefly review our results given in \cite{Chaturvedi:2016rcn} where we have demonstrated that the above expression exactly matches with the large-$c$ limit of the entanglement negativity in  $CFT_{1+1}$ as given in \cite {Calabrese:2012ew,Calabrese:2012nk,Calabrese:2014yza}.

\subsection{Holographic Entanglement Negativity in $AdS_{3}/CFT_{2}$}
In this section we briefly review the application of our conjecture to compute the holographic entanglement negativity for both a pure state described by the $CFT_{1+1}$ vacuum which is dual to a bulk pure $AdS_3$ geometry, and
the finite temperature mixed  state dual to a bulk Euclidean BTZ black hole. 
\subsubsection{Pure $AdS_3$}\label{pu}

In the context of $AdS_3/CFT_2$ correspondence it is well known that the vacuum state of a holographic $CFT_{1+1}$ is dual to pure $AdS_3$ space time whose metric in Poincare coordinates is given below
\begin{equation}
 ds^{2}=\frac{R^2}{z^2}(-dt^2+dz^2+dx^2)\label{puads}.
\end{equation}

where $z$ corresponds to the inverse radial coordinate extending into the bulk, $R$ is the AdS length scale and $(x,t)$ represent the coordinates on the boundary $CFT_{1+1}$.
The length of bulk geodesic ${\cal L}_{\gamma}$  anchored to the subsystem $\gamma$ in the dual $CFT_{1+1}$ in this spacetime is given by \cite{Ryu:2006bv,Ryu:2006ef}
\begin{equation}
 {\cal L}_{\gamma}= 2R ~\ln\big[\frac{l_{\gamma}}{a}\big].\label{geopu}
\end{equation}
The above expression for the length of geodesics which are anchored on various subsystems $\gamma=\{A,B_1,B_2,A\cup B_1, A\cup B_2\}$ as depicted in the fig(\ref{fig:subsystem}), may then be substituted in eq.(\ref{ENCFT4}) to obtain the holographic entanglement negativity  as
\begin{equation}
 {\cal E}=\frac{3R}{4G_N}\ln\big[\frac{l}{a}\big].\label{henpu}
\end{equation}
Note that the contributions from various geodesics in eq.(\ref{ENCFT3}) cancel exactly in the bipartite limit $L\to\infty$ except twice the length of the geodesic anchored to the subsystem-$A$. Hence, upon utilizing the Brown-Hennaux formula $c=\frac{3R}{2G_{N}^{(3)}}$  the above expression for the negativity reduces to
\begin{equation}
{\cal E} =\frac{c}{2}\ln\big[\frac{l}{a}\big]=\frac{3}{2}S_A,\label{hcfpu}
\end{equation}
Remarkably, the above expression exactly matches with the universal part of the replica technique 
result for the $CFT_{1+1}$ vacuum given in eq.(\ref{ENCFTzero}) \cite{Calabrese:2012ew,Calabrese:2012nk}.

\subsubsection{Euclidean BTZ black hole}
In this subsection we review the computation of the holographic entanglement negativity for the bipartite ($A\cup A^c)$ finite temperature mixed state of a $CFT_{1+1}$ which is dual to a bulk Euclidean BTZ black hole \cite{Chaturvedi:2016rcn}. The metric for this Euclidean BTZ black hole is given by
\begin{equation}
 ds^{2}= (r^2-r_h^2)d\tau_{E}^2+ \frac{R^2}{(r^2-r_h^2)} dr^2 +r^2 d\phi^2,\label{BTZm}
\end{equation}
here, $\tau_{E}$ is the compactified Euclidean time $(\tau_{E}\sim \tau_E+\frac{2\pi R}{r_h})$. The coordinate $\phi$ is a periodic for the BTZ black hole i.e $(\phi+2\pi)$ and is uncompactified for the case of BTZ black string. The length of the bulk geodesic  $ {\cal L}_{\gamma}$ that is anchored on the interval $\gamma$ in the boundary $CFT_{1+1}$  is well known in these Euclidean Poincare co-ordinates \cite{Ryu:2006ef} and may be given as follows
\begin{equation}
 {\cal L}_{\gamma}=2R\ln\bigg[\frac{\beta}{\pi a}\sinh[\frac{\pi l_{\gamma}}{\beta}]\bigg],\label{lij}
\end{equation}
 here $a$ is the UV cut-off for the boundary $CFT_{1+1}$,  R is the $AdS_3$ length scale and $l_{\gamma}$ represents the length of the subsystem-$\gamma$. In the $AdS_3/CFT_2$ scenario as shown in fig.(\ref{fig:subsystem}) the geodesic length ${\cal L}_{\gamma}$ given by eq.(\ref{lij}) may be identified for 
 the intervals $\gamma=\{A,B_1,B_2,A\cup B_1, A\cup B_2\}$.
 Using the expression for the geodesic length given by eq.(\ref{lij}) and substituting it in eq.(\ref{ENCFT3}), the holographic entanglement negativity for the finite temperature mixed state of a dual $CFT_{1+1}$ may be obtained as follows  
\begin{equation}
 {\cal E}=\frac{c}{2}\bigg[\ln\bigg\{\frac{\beta}{\pi a}\sinh(\frac{\pi\ell}{\beta})\bigg\}-\frac{\pi \ell}{\beta}\bigg],\label{BTZmatch}
\end{equation}
where we have made use of the previously mentioned Brown-Henneaux formula. Remarkably eq.(\ref{BTZmatch}) obtained from the bulk computation using our conjecture, matches exactly with the large-$c$ limit of  the entanglement negativity for the finite temperature mixed state of  a $CFT_{1+1}$ given by eq.(\ref{larcEN}). The above expression for the holographic entanglement negativity may be concisely expressed as
\begin{equation}
{\cal E}=\frac{3}{2}\bigg[S_A-S^{th}_A\bigg].\label{BTZEN1}
\end{equation}
Here, $S_A$ is the entanglement entropy and $S_{A}^{th}$ is the thermal entropy of the subsystem $A$ for the finite temperature mixed state of a $CFT_{1+1}$. Quite clearly, the above expression  demonstrates that the holographic entanglement negativity obtained from our conjecture captures the distillable quantum entanglement for the bipartite finite temperature mixed state of the dual $CFT_{1+1}$, through the elimination of the thermal contribution.

\section{Holographic entanglement Negativity in $AdS_{d+1}/CFT_{d}$}

In \cite{Chaturvedi:2016rcn} we have proposed that the observations in the previous section 
lead to a higher dimensional extension of our holographic entanglement negativity conjecture for a $CFT_d$ dual to
bulk $AdS_{d+1}$ configurations, in a generic $AdS_{d+1}/CFT_{d}$ scenario. To understand this, it is required to partition the $CFT_d$ into two subsystems $A$ and its complement $A^c$.
Subsequently we consider two other subsystems $B_1$ and $B_2$ adjacent to $A$ and on either either side of it such that $B=(B_1\cup B_2)$. We denote ${\cal A}_{\gamma}$ as the area of the co-dimension two static minimal surface in the bulk $AdS_{d+1}$ geometry, anchored on the subsystems $\gamma$. The holographic entanglement negativity for the bipartite ($A\cup A^c$) quantum state of a $CFT_{d}$ is then given by the following expression
\begin{equation}
{\cal E}=\lim_{B \to A^c}\frac{3}{16G_{N}^{d+1}}\bigg[2 {\cal A}_{A}+{\cal A}_{B_1}+{\cal A}_{B_2}-{\cal A}_{A\cup B_1}-{\cal A}_{A\cup B_2}\bigg]\label{ENCFT5}
\end{equation}
where, $G_{N}^{d+1}$ is the $(d+1)$-dimensional Newton constant and  the bipartite limit $(B \to A^c)$ in 
eq.(\ref{ENCFT5}) corresponds to extending the subsystems $B_1$ and $B_2$ such that $B=(B_1 \cup B_2)$ reduces to the complement $A^c$.
Once again upon making use of the Ryu-Takayanagi conjecture in  eq.(\ref{EEI}), the expression for the holographic negativity in eq.(\ref{ENCFT5}) reduces to the following form
\begin{equation}
 {\cal E}=\lim_{B \to A^c}\frac{3}{4}\bigg[2 S_{A}+S_{B_1}+S_{B_2}-S_{A\cup B_1}-S_{A\cup B_2}\bigg],\label{ENCFTSAd}
\end{equation}
Re-expressing the above expression as the sum of holographic mutual informations ${\cal I}(A,B_i)$, we obtain
 \begin{eqnarray}
 {\cal E}&=&\lim_{B\to A^{c}} \frac{3}{4}\big[{\cal I}(A,B_1)+{\cal I}(A,B_2)\big]\label{ENMII}
  \end{eqnarray}
where, the holographic mutual information ${\cal I}(A,B_i)$ ($i=1,2$)  are given as follows
 \begin{eqnarray}
  {\cal I}(A,B_i)&=&S_{A}+S_{B_i}-S_{A\cup B_i},\nonumber\\
&=&\frac{1}{4G_{N}^{(d+1)}}({\cal A}_{A}+{ \cal A}_{B_i}-{\cal A}_{A\cup B_i}).
 \end{eqnarray}
In the following subsections, using the above mentioned holographic conjecture we will obtain the entanglement negativity for both a pure state described by the $CFT_d$ vacuum which is dual to the bulk pure $AdS_{d+1}$ space time and the finite temperature mixed state dual to a bulk $AdS_{d+1}$-Schwarzschild black hole. It will be demonstrated that the holographic entanglement negativity for both of these examples, exhibits certain universal features that are independent of the dimensionality of the conformal field theory. As mentioned in the Introduction this serves as a strong consistency check for the higher dimensional extension of our holographic conjecture although a bulk proof along the lines of \cite{Lewkowycz:2013nqa}  is an outstanding open issue which needs to be addressed.

\subsection{ Pure vacuum state of a CFT$_d$ dual to pure AdS$_{d+1}$}
In this section we employ our  conjecture in the $AdS_{d+1}/CFT_d$ scenario, to compute the holographic entanglement negativity for a bipartite pure state described by the $CFT_d$ vacuum which is dual to the pure $AdS_{d+1}$ spacetime. We consider the partitioning of the $CFT_d$ into the subsystem $A$ of rectangular strip geometry and its complement $A^c$. We then consider two other finite subsystems $B_1$ and $B_2$ of rectangular strip geometries adjacent to the subsystem $A$ and on either either side of it, such that $B=(B_1\cup B_2)$.  The metric of pure $AdS_{d+1}$ space time in Poincare coordinates is given by
\begin{equation}
ds^2=\frac{1}{z^2}\Big(-dt^2+\sum_{i=1}^{d-1}(dx^i)^2+dz^2\Big),\label{Poads}
\end{equation}
where $z$ is the inverse radial coordinate and $(x^i,t)$ are the coordinates on the boundary $CFT_d$($i=1,2...,d-1$). Note that the $AdS$ length scale has been set to unity. We consider the subsystem $A$ to be a rectangular strip with the following dimensions
$x^1 \equiv [-\frac{l}{2},\frac{l}{2}]~~x^k=[-\frac{L}{2},\frac{L}{2}],~~k=2,...,(d-1)$ and the rest of the system is denoted as $A^c$. In analogy with the $AdS_3/CFT_2$ scenario we consider two large but finite subsystems $B_1$ and $B_2$ adjacent to the subsystem $A$, defined by the coordinates $x^1\in\big[-L,-\frac{\ell}{2}]$, $ x^k\in \big[\frac{-L_2}{2},\frac{L_2}{2}]$ and $x^1\in\big[\frac{\ell}{2},L]$, $ x^k\in \big[\frac{-L_2}{2},\frac{L_2}{2}]$ respectively. In order to determine the area of the required  bulk static minimal surfaces anchored to the boundary subsystem, the following area functional has to be extremized  \cite {Ryu:2006ef}.
\begin{equation}
{\cal A}_{\gamma}= L_2^{d-2}\int_{x_j^{1}}^{x_i^{1}} dx^1 \frac{\sqrt{1+(\frac{dz}{dx^1})^2}}{z^{d-1}}.\label{intAg}
\end{equation}
The Euler-Lagrange equation for the extremization of the above area functional is then given as
\begin{equation}\label{zst}
\frac{dz}{dx^1}=\frac{\sqrt{z_{*}^{2(d-1)}-z^{2(d-1)}}}{z^{d-1}},
\end{equation}
where $z=z_*$ is the turning point of the  minimal surface. 
The areas of  minimal surfaces ${\cal A}_{A},{\cal A}_{B_1}$ and ${\cal A}_{A\cup B_1}$ may then be obtained through the integral given in eq.(\ref{intAg}) and eq.(\ref{zst}) as described in \cite{Ryu:2006ef}
\begin{eqnarray}
 {\cal A}_{A}&=\frac{2}{d-2}\Big(\frac{L_2}{a}\Big)^{d-2}-s_0~\Big(\frac{L_2}{l}\Big)^{d-2}\label{Aapu}\\
 {\cal A}_{B_1}&=\frac{2}{d-2}\Big(\frac{L_2}{a}\Big)^{d-2}-s_0~\Big(\frac{L_2}{L-\frac{l}{2}}\Big)^{d-2}\label{Ab1pu}\\
  {\cal A}_{A\cup B_1}&=\frac{2}{d-2}\Big(\frac{L_2}{a}\Big)^{d-2}-s_0~\Big(\frac{L_2}{L+\frac{l}{2}}\Big)^{d-2},\label{Aab1pu}
\end{eqnarray}
where, $s_0$ is a constant given as follows
\begin{equation}\label{s0}
s_0=\frac{2^{d-1}\pi^{(d-1)/2}}{d-2}\Bigg(\frac{\Gamma(\frac{d}{2(d-1)})}{\Gamma(\frac{1}{2(d-1)})}\Bigg)^{d-1}.
\end{equation}
 Note that the subsystem $A$ has been chosen to be symmetric along the partitioning direction leading to the equality of the  minimal areas  ${\cal A}_{B_1}={\cal A}_{B_2}$ and ${\cal A}_{A\cup B_1}={\cal A}_{A\cup B_2}$. This identification reduces the expression given in eq.(\ref{ENCFT5}), for the holographic entanglement negativity to the following form   
\begin{equation}\label{ken}
{\cal E} = \lim_{B\rightarrow A^c}\frac{3}{8G_{N}^{d+1}} \bigg[{\cal A}_{A}+{\cal A}_{B_1}-{\cal A}_{A\cup B_1}\bigg].
\end{equation} 
Having obtained the required expressions for the areas of  minimal surfaces given by equations (\ref{Aapu}), (\ref{Ab1pu}) and (\ref{Aab1pu}), we may now utilize eq.(\ref{ken}) to determine the holographic entanglement negativity to be
\begin{equation}
  {\cal E} = \lim_{L\to\infty}\frac{3}{8G_{N}^{d+1}} \bigg[\frac{2}{d-2}\Big(\frac{L_2}{a}\Big)^{d-2}-s_0\bigg\{\Big(\frac{L_2}{l}\Big)^{d-2}+\Big(\frac{L_2}{L-\frac{l}{2}}\Big)^{d-2}-\Big(\frac{L_2}{L+\frac{l}{2}}\Big)^{d-2}\bigg\} \bigg]
\end{equation}
This leads us to the following expression
\begin{eqnarray}
 {\cal E} =\frac{3}{8G_{N}^{d+1}}\bigg[\frac{2}{d-2}\Big(\frac{L_2}{a}\Big)^{d-2}-s_0\Big(\frac{L_2}{l}\Big)^{d-2}]
\end{eqnarray}
Quite interestingly, upon utilizing the Ryu-Takayanagi conjecture given in eq.(\ref{EEI}) the above expression for holographic entanglement negativity of the pure vacuum state of the $CFT_d$  reduces to the following form
\begin{equation}
  {\cal E}=\frac{3}{2}[S_A]
\end{equation}
Remarkably, this result is identical in form to entanglement negativity for the pure state described by  the $CFT_{1+1}$ vacuum, as given in eq.(\ref{hcfpu}) for the corresponding $AdS_3/CFT_2$ example.  Hence, this result serves as a first consistency check for the higher dimensional extension of our holographic conjecture proposed in \cite{Chaturvedi:2016rcn}.

\subsection{ Finite temperature mixed state of a CFT$_d$ dual to AdS$_{d+1}$ Schwarzschild black hole}
\begin{figure}[ht!]
\centering
\includegraphics[width =3in,height=2.2in]{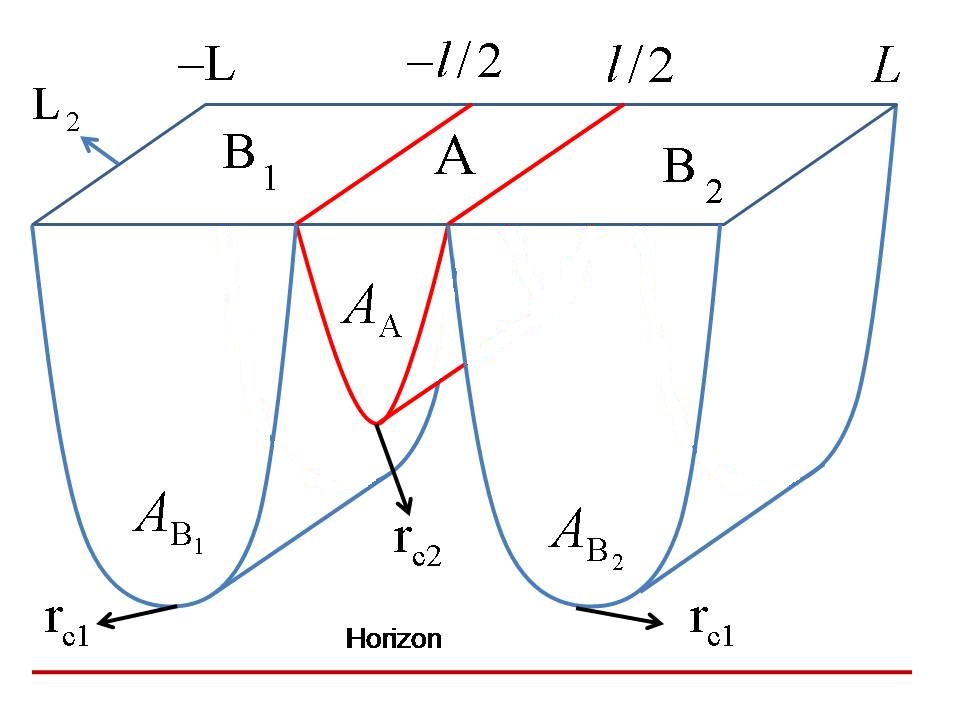}
\caption{\label{fig:AdS41} Schematic of static minimal surfaces anchored on the subsystems $A$, $B_1$ and $B_2$  in the low temperature regime.}\end{figure}

 In this section we apply our holographic conjecture to another higher dimensional example in the
 $AdS_{d+1}/CFT_d$ scenario. In this context, we compute the holographic entanglement negativity for a bipartite finite temperature mixed state of a holographic $CFT_d$ dual to a bulk $AdS_{d+1}$-Schwarzschild black hole. In this case, the $CFT_d$ is partitioned into the subsystem $A$ of rectangular strip geometry and its complement $A^c$. Once again we consider two finite subsystems $B_1$ and $B_2$ of rectangular strip geometries adjacent to the subsystem $A$ and on either either side of it, such that $B=(B_1\cup B_2)$ as shown schematically in the fig.(\ref{fig:AdS41}). The metric for a 
 $AdS_{d+1}$-Schwarzschild black hole with a planar horizon in the Poincare coordinates is given by
 \begin{equation}
  ds^2= -r^2(1-\frac{r_h^d}{r^d})dt^2+\frac{dr^2}{r^2(1-\frac{r_h^d}{r^d})}+r^2d\vec{x}^2,
 \end{equation}
 where $r_h$ is the horizon radius of the black hole with the Hawking temperature $T=r_h d/4\pi$ and $ \vec{x}\equiv(x,x^i) $ are the spatial co-ordinates on the boundary and $i=1..(d-2)$. Here we set the AdS length scale $R$ to unity. The holographic entanglement negativity in this case is given by the eq.(\ref{ENCFT5}) in terms of the areas of the bulk co dimension two static minimal surfaces anchored on the corresponding subsystems (see fig.(\ref{fig:AdS41})).
As is evident from  fig.(\ref{fig:AdS41}) the subsystem $A$ corresponds to a spatial region on the $d$-dimensional boundary defined by the coordinates $x\in\big[-\frac{\ell}{2},\frac{\ell}{2}]$, $x^i\in \big[\frac{-L_2}{2},\frac{L_2}{2}]$ where $L_2>>\ell$. Similarly, the spatial region describing the subsystems $B_1$ and $B_2$ are defined by the coordinates $x\in\big[-L,-\frac{\ell}{2}]$, $ x^i\in \big[\frac{-L_2}{2},\frac{L_2}{2}]$ and $x\in\big[\frac{\ell}{2},L]$, $ x^i\in \big[\frac{-L_2}{2},\frac{L_2}{2}]$ respectively such that $L>>\ell$. Note that from the above the spatial region corresponding to the subsystem $ A\cup B_1$ is defined by the coordinates $ x\in\big[-L,\frac{\ell}{2}]$, $x^i\in \big[\frac{-L_2}{2},\frac{L_2}{2}]$. 

\begin{figure}[ht!]
\centering
\includegraphics[width =3in,height=2.2in]{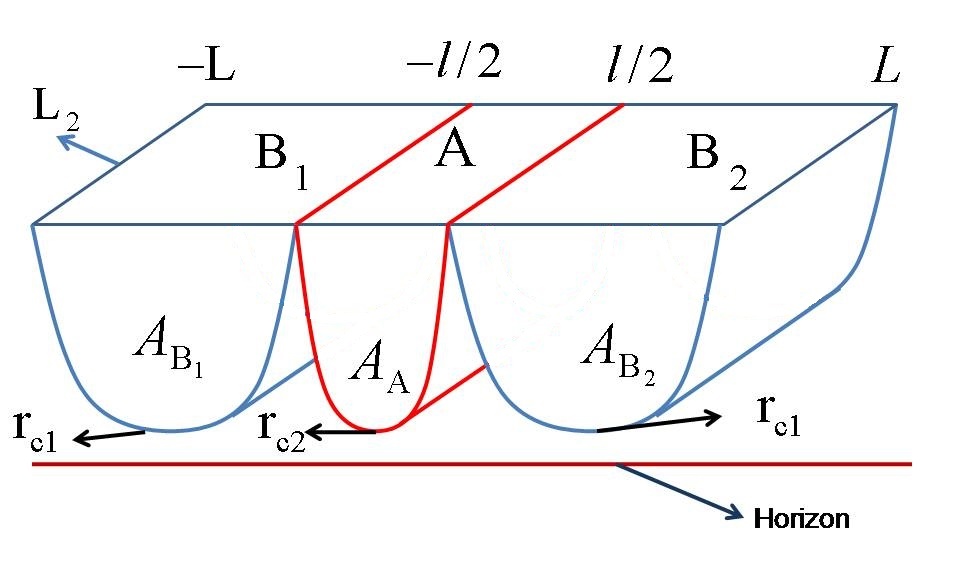}
\caption{\label{fig:AdS42}Schematic of static minimal surfaces anchored on the subsystems $A$, $B_1$ and $B_2$ in the high temperature regime.}\end{figure}

Notice that the subsystem $A$ has been chosen to be symmetric along the partitioning direction as shown in the fig.(\ref{fig:AdS41}). This leads to the equality of the minimal areas  ${\cal A}_{B_1}={\cal A}_{B_2}$ and ${\cal A}_{A\cup B_1}={\cal A}_{A\cup B_2}$. This identification reduces the expression for the holographic entanglement negativity in eq.(\ref{ENCFT5}), to the following form   
\begin{equation}\label{ke}
{\cal E} = \lim_{B\rightarrow A^c}\frac{3}{8G_{N}^{(d+1)}} \bigg[{\cal A}_{A}+{\cal A}_{B_1}-{\cal A}_{A\cup B_1}\bigg].
\end{equation}
The expression for the area of the surface which is anchored to a subsystem in the  $CFT_d$ dual to a bulk planar $AdS_{d+1}$-Schwarzschild  black hole is given in \cite{Fischler:2012ca} as 
\begin{equation}
 {\cal A}=L_{2}^{d-2}\int dr r^{d-2}\sqrt{r^2x'^2+\frac{1}{r^2(1-\frac{r_h^d}{r^d})}}.\label{gena}
\end{equation}
Extremizing the above area integral leads to the following Euler-Lagrange equation
\begin{equation}
  \int_{x_{1}}^{x_{2}} dx=\frac{2}{ r_{c} } \int^1_{0} \frac{u^{d-1} du}{\sqrt{(1-u^{2d-2})} }(1-\frac{r_h^d}{r_{c}^d}u^d)^{-\frac{1}{2}},\label{tur}
\end{equation}
here, $x_{1}$ and $x_{2}$ represent the end point of the subsystem under consideration, $r_c$ represents the turning point of the static minimal surface and the integration variable is given by $u=\frac{r_c}{r}$. After integration, the resulting equation may be inverted to obtain the turning radius $r_c$. This may then be substituted in the expression for the area of the  minimal surface. The area integral in eq.(\ref{gena}) written in terms of the variable $u$ may be expressed as
\begin{equation}
{\cal A} = 2 L_{2}^{d-2}r_{c}^{d-2} \int^1_{0} \frac{du}{u^{d-1}\sqrt{(1-u^{2d-2})} }(1-\frac{r_h^d}{r_{c}^d}u^d)^{-\frac{1}{2}}.\label{ar}
\end{equation}
The integrals in eq.(\ref{tur}) and eq.(\ref{ar}) are not analytically solvable. Therefore to compute these integrals we adopt the method developed in \cite{Fischler:2012ca} where the authors employ a certain  expansion technique in terms of Gamma functions to compute these integrals order by order. Denoting the turning points of the static minimal surfaces whose areas are given as ${\cal A}_{B_1},{\cal A}_{A}$ and 
${\cal A}_{A\cup B_1}$ to be $r_{c1}$, $r_{c2}$ and $r_{c3}$ respectively, it is possible to obtain the expression for the subsystem lengths using eq. (\ref {tur}) as follows \cite{Fischler:2012ca} 
 \begin{eqnarray}
L-\frac{\ell}{2}&=&\frac{2}{ r_{c1} } \sum_{n=0}^{\infty}g_n \bigg(\frac{r_h}{r_{c1}}\bigg)^{nd},\label{rc1}\\
\ell&=&\frac{2}{ r_{c2} } \sum_{n=0}^{\infty} g_n\bigg(\frac{r_h}{r_{c2}}\bigg)^{nd},\label{rc2}\\
L+\frac{\ell}{2}&=&\frac{2}{ r_{c3} } \sum_{n=0}^{\infty} g_n\bigg(\frac{r_h}{r_{c3}}\bigg)^{nd}.\label{rc3}\\
\end{eqnarray}
Here $g_n$ is given by
\begin{equation}
 g_n=(\frac{1}{dn+1}) \frac{\Gamma(n+\frac{1}{2})}{\Gamma(n+1)}\frac{\Gamma(\frac{d(n+1)}{2(d-1)})}{\Gamma(\frac{(dn+1)}{2(d-1)})}.\label{bn}
\end{equation}
The expressions for the minimal surfaces ${\cal A}_{B_1},{\cal A}_{A}$ and ${\cal A}_{A\cup B_1}$ may be expressed as
\begin{equation}
 {\cal A}_{B_1} =\frac{2}{d-2}(\frac{L_{2}}{a})^{d-2}+ 2 L_{2}^{d-2}r_{c1}^{d-2} \sum_{n=0}^{\infty} a_n\bigg(\frac{r_h}{r_{c1}}\bigg)^{nd},~~~~~~\label{a12}
\end{equation}
  \begin{equation}
 {\cal A}_{A} =\frac{2}{d-2}(\frac{L_2}{a})^{d-2} + 2 L_{2}^{d-2}r_{c2}^{d-2} \sum_{n=0}^{\infty} a_n(\frac{r_h}{r_{c2}})^{nd},~~~~~~\label{a23}
 \end{equation}
and
 \begin{equation}
 \begin{split}
  {\cal A}_{A\cup B_1} = \frac{2}{d-2}(\frac{L_{2}}{a})^{d-2}+ 2 L_{2}^{d-2}r_{c3}^{d-2} \sum_{n=0}^{\infty} a_n \bigg(\frac{r_h}{r_{c3}}\bigg)^{nd}.~~~~~~\label{a13}
 \end{split} 
 \end{equation}
Here $a_n$ is given by
\begin{equation}
 a_n=\frac{1}{2(d-1)} \frac{\Gamma(n+\frac{1}{2})}{\Gamma(n+1)}\frac{\Gamma(\frac{d(n-1)+2}{2(d-1)})}{\Gamma(\frac{(dn+1)}{2(d-1)})}\label{sn}
\end{equation}
It is to be noted that the integral for the area in eq.(\ref{ar}) is divergent and has to be regulated by an infrared cut-off of the bulk (say $r_{in}$) which is related to the UV cut-off $(a)$ of the $d$-dimensional boundary CFT as $ r_{in}=1/a$ \cite{Fischler:2012ca}. Having performed all the integrals we substitute eq.(\ref{a12}), eq.(\ref{a13}) and eq.(\ref{a23}) in eq.(\ref{ke}) to arrive at the expression for the holographic entanglement negativity as
\begin{equation}
\begin{split}
{\cal E} = \lim_{L \rightarrow \infty}\frac{3}{8G_{N}^{(d+1)}} \bigg[&\frac{2}{d-2}(\frac{L_{2}}{a})^{d-2}+ 2 L_{2}^{d-2}r_{c1}^{d-2} \sum_{n=0}^{\infty}a_n \big(\frac{r_h}{r_{c1}}\big)^{nd}\\+&2 L_{2}^{d-2}r_{c2}^{d-2} \sum_{n=0}^{\infty}  a_n\big(\frac{r_h}{r_{c2}}\big)^{nd}
 -2 L_{2}^{d-2}r_{c3}^{d-2} \sum_{n=0}^{\infty} a_n\big(\frac{r_h}{r_{c3}}\big)^{nd}  \bigg].
 \end{split}
\end{equation}
Notice that it is required to invert the expressions in eq.(\ref{rc1}), 
eq.(\ref{rc2}) and eq.({\ref{rc3}) to obtain $r_{c1}$, $r_{c2}$, $r_{c3}$ and then substitute those in the above equation to obtain the holographic negativity as a function of the temperature and the length $(\ell)$ of the subsystem $A$.

\subsection{Low temperature regime}
In this section, we compute the holographic entanglement negativity for the bipartite finite temperature mixed state of the $CFT_d$ in the low temperature regime. This regime corresponds to the temperature $T\ell<<1$, which in the bulk translates to the case where the horizon is at a large distance from the turning point $r_{c2}$ of the static  minimal surface anchored on the subsystem $A$. This is equivalent to the condition $r_{c2}>>r_h $ as shown in the fig.(\ref{fig:AdS41}). As $r_h\ell<<1$, the expression for the turning point $r_{c2}$ may be obtained perturbatively employing the technique described in \cite{Fischler:2012ca} as follows
 \begin{equation}
 r_{c2}= \frac{b_0}{\ell} \bigg[1+b_1 (r_h\ell)^d+O[r_h^{2d}\ell^{2d}]\bigg],\label{rc2l}
\end{equation}
where $b_0$, $b_1$ are constants given by
\begin{equation}
 b_0= \frac{2 \sqrt{\pi} \Gamma(\frac{d}{2(d-1)})}{\Gamma(\frac{1}{2(d-1)})},
\end{equation}
\begin{equation}
 b_1=  \frac{1}{2(d+1)}\frac{2^{\frac{1}{d-1}-d}\Gamma(1+\frac{1}{2(d-1)})\Gamma(\frac{1}{2(d-1)})^{d+1}}{\pi^{\frac{d+1}{2}}\Gamma(\frac{1}{2}+\frac{1}{(d-1)})\Gamma(\frac{d}{2(d-1)})^{d}}.
\end{equation}
We find the area ${\cal A}_{A}$ by substituting the expression for $r_{c2}$ given by eq.(\ref{rc2l}) in the eq.(\ref{a23}) while keeping only the leading terms in $(r_h\ell)^d$ as follows     
\begin{equation}\label{a23l}
\begin{split}
 {\cal A}_{A} = \frac{2}{d-2}(\frac{L_{2}}{a})^{d-2}+ s_0(\frac{L_2}{\ell})^{d-2}\bigg[1+s_1 (r_h\ell)^d+O[ (r_h\ell)^{2d} ]\bigg],
\end{split}
\end{equation}
where $s_0$ and $s_1$ are given by
\begin{eqnarray}
 s_0=\frac{2^{d-2}\pi^{\frac{d-1}{2}} \Gamma(-\frac{d-2}{2(d-1)})\Gamma(\frac{d}{2(d-1)})^{d-2}}{(d-1)\Gamma(\frac{1}{2(d-1)})^{d-1}},
\end{eqnarray}
\begin{equation}
 s_1=\frac{\Gamma(\frac{1}{2(d-1)})^{d+1}}{2^{d+1}\pi^{\frac{d}{2}}\Gamma(\frac{d}{2(d-1)})^d\Gamma(\frac{d+1}{2(d-1)})}\bigg(\frac{\Gamma(\frac{1}{d-1})}{\Gamma(-\frac{d-2}{2(d-1)})}+\frac{2^\frac{1}{d-1}(d-2)\Gamma(1+\frac{1}{2(d-1)})}{\sqrt{\pi}(d+1)}\bigg).
\end{equation}
The subsystems $B_1$ and $A\cup B_1$ in the boundary $CFT_d$ with lengths 
$(L-\ell/2)$ and $(L+\ell/2)$ ~along the $x$ direction are very large in the limit $B\to A^c$  $(L \to \infty)$. Therefore, the  minimal surfaces described by the areas ${\cal A}_{B_1}$ and  ${\cal A}_{A\cup B_1}$ will extend deep into the bulk approaching the black hole horizon even at low temperatures i.e., ($r_{c1}\sim r_h)$ and ($r_{c3}\sim r_h)$. Hence, in order to compute the expressions for the areas ${\cal A}_{B_1}$  and  ${\cal A}_{A\cup B_1}$ we employ the method developed by the authors in \cite{Fischler:2012ca} for the case when the minimal surfaces approach the black hole horizon as described earlier. Through this procedure we obtain the expression for the turning point $r_{c1}$ for the  minimal surface anchored on the subsystem $B_1$ 
as follows
\begin{equation}
 r_{c1}= r_h(1+\epsilon_1),\label{rc1l}
\end{equation}
\begin{equation}
\epsilon_1= k_2 \ e^{-\sqrt{\frac{d(d-1)}{2}} r_h (L-\frac{\ell}{2})} ,\label{eps}
\end{equation}
where $k_2$ is a constant given by 
\begin{eqnarray}\label{k2}
 k_2 = \frac{1}{d} e^{\sqrt{\frac{d(d-1)}{2}}c_1}, 
 \end{eqnarray} 
\begin{equation}
 c_1=\frac{2\sqrt{\pi}\Gamma(\frac{d}{2(d-1)})}{\Gamma(\frac{1}{(d-1)})}+ 
 \sum_{n=1}^\infty\bigg(\frac{2}{(1+nd)} \frac{\Gamma(n+\frac{1}{2})}{\Gamma(n+1)}\frac{\Gamma(\frac{d(n+1)}{2(d-1)})}{\Gamma(\frac{dn+1}{2(d-1)})}-\frac{\sqrt{2}}{\sqrt{d(d-1)}n}\bigg).
\end{equation}
Substituting the expressions given by eq.(\ref{rc1l}) and eq.(\ref{eps}) in eq.(\ref{a12})  we obtain the area ${\cal A}_{B_1}$ as an expansion in $\epsilon_1$ up to $O[\epsilon_1]$ as
\begin{equation}\label{a12l}
 {\cal A}_{B_1} = \frac{2}{d-2}(\frac{L_{2}}{a})^{d-2}+\bigg[  L_2^{d-2} r_h^{d-1} (L-\frac{\ell}{2})+ L_2^{d-2} r_h^{d-2}(k_1- \sqrt{\frac{2(d-1)}{d}}\epsilon_1  )+O[\epsilon_1^2]\bigg] ,
\end{equation}
where, $k_1$ is a constant defined as
\begin{equation}
\begin{split}
 k_1=&2\bigg[-\frac{\sqrt{\pi}(d-1)\Gamma(\frac{d}{2(d-1)})}{(d-2)\Gamma(\frac{1}{2(d-1)}}\\&+\sum_{n=1}^{\infty}\frac{1}{1+nd} (\frac{d-1}{d(n-1)+2}) \frac{\Gamma(n+\frac{1}{2})}{\Gamma(n+1)}\frac{\Gamma(\frac{d(n+1)}{2(d-1)})}{\Gamma(\frac{(dn+1)}{2(d-1)})}\bigg].
\end{split}
\end{equation}
Repeating the above procedure we find the expressions for  $r_{c3}$ and ${\cal A}_{A\cup B_1}$ from eq.(\ref{rc3}) and eq.(\ref{a13}) as follows
\begin{equation}\label{rc3l}
 r_{c3}= r_h(1+\epsilon_3),
\end{equation}
\begin{equation}
\epsilon_3= k_2 \ e^{-\sqrt{\frac{d(d-1)}{2}} r_h (L+\frac{\ell}{2})}, 
\end{equation}
\begin{equation}\label{a13l}
\begin{split}
 {\cal A}_{A\cup B_1} = \frac{2}{d-2}(\frac{L_{2}}{a})^{d-2}+ \bigg[& L_2^{d-2} r_h^{d-1} (L+\frac{\ell}{2})\\&+ L_2^{d-2} r_h^{d-2}(k_1- \sqrt{\frac{2(d-1)}{d}}\epsilon_3  )+O[\epsilon_3^2] \bigg].
\end{split}
\end{equation}
Now we substitute the expressions given by eq.(\ref{a12l}), eq.(\ref{a23l})and eq.(\ref{a13l})
for the areas of minimal surfaces ${\cal A}_{B_1},{\cal A}_{A}$ and ${\cal A}_{A\cup B_1}$ obtained in the low temperature regime, in eq.(\ref{ke}). This leads to the following expression for the entanglement negativity 
${\cal E}$ in the low temperature regime as
\begin{equation}
\begin{split}
 {\cal E}= \frac{3}{8G_N^{(d+1)}}\bigg[\frac{2}{d-2}(\frac{L_{2}}{a})^{d-2}+s_0(\frac{L_2}{\ell})^{d-2}[1+ s_1 (r_h\ell)^d]  -Vr_h^{d-1}\bigg],
 \end{split}
\end{equation}
where $V=\ell L_2^{d-2} $ is the $(d-1)$-dimensional volume of the subsystem-$A$. The above expression for the holographic entanglement negativity in the low temperature regime may be re expressed in a concise form as
\begin{equation}
 {\cal E}=\frac{3}{2}\bigg[ S_A-S_A^{th}\bigg].
\end{equation}
In the above expression $S_A$ is the entanglement entropy for the subsystem $A$ of rectangular strip geometry for the finite temperature mixed state of a $CFT_d$ dual to a $AdS_{d+1}$-Schwarzschild black hole and $S_A^{th}=\frac{V r_h^{d-1}}{4G_N^{(d+1)}}$ represents the thermal entropy of the subsystem-$A$. Remarkably, from the above equation we observe that the entanglement negativity captures the distillable quantum entanglement through the removal of the thermal contribution in this regime and is identical in form to the corresponding $AdS_3/CFT_2$ result. This is very significant as our conjecture reproduces the universal feature of the entanglement negativity for the finite temperature mixed state of a holographic $CFT_{1+1}$, in higher dimensions. Naturally this provides a strong consistency check for the higher dimensional extension of our holographic negativity conjecture for the low temperature regime in the $AdS_{d+1}/CFT_d$ scenario.
We now extend the above analysis to the high temperature regime in the next subsection.

\subsection{High temperature regime}

At high temperatures, the turning point $r_{c2}$ of the  minimal surface with the area ${\cal A}_{A}$ approaches close to the black hole horizon which is described by the condition $r_{c2}\sim r_h$ as shown in fig.(\ref{fig:AdS42}). Note that the high temperature regime also implies a large horizon radius $(r_h)$ for the bulk $AdS_{d+1}$-Schwarzschild black hole. Following \cite{Fischler:2012ca} we obtain  ${\cal A}_{A}$
in a near horizon expansion in $\epsilon_2$ up to $O[\epsilon_2]$ by considering $r_{c2}= r_h(1+\epsilon_2)$  as follows
\begin{equation}\label{rc2h}
 r_{c2}= r_h(1+\epsilon_2),
\end{equation}
\begin{equation}
 \epsilon_2= k_2 \ e^{-\sqrt{\frac{d(d-1)}{2}}r_h\ell} ,
\end{equation}

\begin{equation}\label{a23h}
 {\cal A}_{A} = \frac{2}{d-2}(\frac{L_{2}}{a})^{d-2}+ \bigg[ L_2^{d-2} r_h^{d-1} (\ell)+ L_2^{d-2} r_h^{d-2}(k_1- \sqrt{\frac{2(d-1)}{d}}\epsilon_2  ) +O[\epsilon_2^2] \bigg].
\end{equation}

We now turn to the evaluation of the other two  minimal surfaces described by the areas ${\cal A}_{B_1}$ and  ${\cal A}_{A\cup B_1}$. Note that as described earlier these surfaces always probe the near horizon regime
both at low and at high temperatures due to the limit $B\to A^c$ or equivalently $L\to \infty$. Hence we may use the general expression for these  minimal areas given in eq.(\ref{a12l}) and eq.(\ref{a13l}) in the high temperature regime as well. Following this we substitute the areas of all the three  minimal surfaces given by  eq.(\ref{a23h}), eq.(\ref{a12l}) and eq.(\ref{a13l}) in the expression for the holographic entanglement negativity given by eq(\ref{ke}). This leads us to the expression for the holographic entanglement negativity in the high temperature regime as follows
\begin{equation}
{\cal E} =\frac{3}{8G_N}\bigg[  \frac{2}{d-2}(\frac{L_{2}}{a})^{d-2}+L_2^{d-2} r_h^{d-2}(k_1- \sqrt{\frac{2(d-1)}{d}}k_2 \ e^{-\sqrt{\frac{d(d-1)}{2}}r_h(\ell)} )\bigg].\label{al}
\end{equation}
Observe that as earlier for the low temperature regime we may re express the above equation in the high temperature regime also in the following concise form 
\begin{equation}
{\cal E} =\frac{3}{2}\bigg[ S_A-S_A^{th}\bigg].\label{ENSch1}
\end{equation}
From the above expression notice that as earlier for the low temperature regime, the entanglement negativity for the high temperature regime also leads to the distillable quantum entanglement through the removal of the thermal contribution. Significantly, we once again observe that the above expression is identical in form to the corresponding $AdS_3/CFT_2$ result given in eq.(\ref{BTZEN1}). Hence, in the high temperature regime also our conjecture reproduces the universal feature of the entanglement negativity for the finite temperature mixed state of a holographic $CFT_{1+1}$, in higher dimensions. Clearly, the results of the last two sections serve as strong consistency checks for the universality of our conjecture and its relevance to $d$-dimensional CFTs in a generic $AdS_{d+1}/CFT_d$ scenario.

\section{Summary and Conclusions}

To summarize, in this article we have examined the consistency of the higher dimensional $AdS_{d+1}/CFT_d$ extension of our holographic entanglement negativity conjecture proposed in the $AdS_3/CFT_2$ context \cite{Chaturvedi:2016rcn} (CMS), through the application to specific examples. In this connection, utilizing the higher dimensional $AdS_{d+1}/CFT_{d}$ extension of our conjecture we have computed the holographic entanglement negativity for bipartite pure and finite temperature mixed states of dual $CFT_d$s. These include the bipartite pure state of the $CFT_d$ vacuum dual to a bulk pure $AdS_{d+1}$ geometry and the finite temperature mixed state dual to a $AdS_{d+1}$-Schwarzschild black hole.
We have demonstrated that holographic entanglement negativity for the pure vacuum state
is proportional to the holographic entanglement entropy. Very significantly the expression for the holographic entanglement negativity is identical in form ( same proportionality constant) to the corresponding case of the pure vacuum state in a holographic $CFT_{1+1}$ \cite{Chaturvedi:2016rcn}. Furthermore, the holographic entanglement negativity for the finite temperature mixed state in question computed from our conjecture correctly leads to the distillable entanglement through the elimination of the thermal contribution. Significantly, once again this is identical in form to the $AdS_3/CFT_2$ result
\cite{Chaturvedi:2016rcn}. Interestingly, our results exactly reproduce (in form) the universal features of the entanglement negativity of $CFT_{1+1}$ in higher dimensions and hence, constitute very strong consistency check for the higher dimensional extension of our conjecture despite a bulk proof along the lines of \cite{Lewkowycz:2013nqa} being a significant open issue which needs attention.

It is well known that mixed state entanglement has significant implications for understanding diverse fields including 
quantum information theory, condensed matter physics and issues of quantum gravity such as black hole formation and collapse and the information loss paradox. As described earlier, the entanglement negativity serves as a measure to characterize such mixed state entanglement. Hence, we expect that our entanglement negativity conjecture  
for holographic conformal field theories to lead to wide ramifications in disparate fields. For example entanglement negativity is related to the topological order and topological entanglement in diverse condensed matter systems described by conformal field theories. Furthermore, entanglement negativity is also expected to have significant import for
the investigation of high temperature superconductivity, quantum phase transitions, quantum quenches and thermalization which involve {\it entanglement evolution}. In particular our conjecture should be significant in studying strongly coupled many body systems in the context of the $AdS$ condensed matter theory ($AdS/CMT$) correspondence. It is also well known that entanglement entropy and mutual information have played an important role in the investigation of the information loss paradox and the associated black hole {\it firewall problem}. Interestingly, our conjecture directly relates the holographic entanglement negativity and the associated distillable quantum entanglement with the holographic mutual information. Naturally, this indicates that our conjecture ( or a covariant version thereof) should also have crucial implications for the study of the Information Loss Paradox and the black hole {\it firewall problem}. We hope to return to these interesting issues in the near future.

\section {Acknowledgement } All of us would like to thank Ashoke Sen for crucial discussions and suggestions.
We would also like to thank Sayantani Bhattacharyya and Saikat Ghosh for extremely useful discussions and insights. The work of Pankaj Chaturvedi is supported by  Grant No. 09/092(0846)/2012-EMR-I, from the Council of Scientific and Industrial Research (CSIR), India.

\begin{appendices}

\section{Review of entanglement negativity in $CFT_{1+1}$}\label{FTEN}

In this appendix,  we review the procedure for obtaining the entanglement negativity in a $CFT_{1+1}$ described by the authors Calabrese et al. in \cite{Calabrese:2014yza}. As discussed in the introduction, the entanglement negativity of a mixed described by the bipartite system consisting of subsystems $A_1$ and $A_2$ ($A=A_1\cup A_2$) embedded in a larger tripartite system $A_1\cup A_2\cup A^c$ may be given as
\begin{equation}
{\cal E} = \log \big(Tr\mid\rho_{A}^{T_{2}}\mid\big),\label{ENegdef}
\end{equation}
here, $\rho_{A}=Tr_{A^c}\left(\rho\right)$ is reduced density matrix and the superscript $T_2$ represents the operation of the partial transpose on this reduced density matrix $\rho_{A}^{T_{2}}$ as described in eq.(\ref{trace}). 

Note that for extended quantum many body systems like quantum field theories just as for entanglement entropy the computation of the entanglement negativity involves an infinite dimensional density matrix. Hence, the application of the above formula for the entanglement negativity becomes problematic. However, for  this issue may be addressed in the framework of the {\it replica technique} proposed in 
\cite{Calabrese:2014yza} mentioned earlier. Using this technique the authors were able to compute the entanglement negativity for bipartite quantum states of a $CFT_{1+1}$, by relating it to the quantity $Tr(\rho_{A}^{T_{2}})^n $. From the computation of the entanglement entropy it is well known that the quantity $Tr(\rho_A)^n$ is given by the following four point twist correlator
\begin{equation}
Tr(\rho_A)^{n}=\big<{\cal T}_{n}(u_1)\overline{{\cal T}}_{n}(v_1){{\cal T}}_{n}(u_2)\overline{\cal T}_{n}(v_2)\big>.\label{rhoAnT}
\end{equation}
In this regard, the operation of the partial transpose $(\rho_A^{T_2})$ of the reduced density matrix $\rho_A$ has the effect of exchanging upper and lower edges of the branch cut along the interval $A_2$ on a $n_e$-sheeted Riemann surface. Thus the quantity $Tr(\rho_A^{T_2})^n$ may be expressed in terms of a four point twist correlator as 
\begin{equation}
Tr(\rho_A^{T_2})^{n}=\big<{\cal T}_{n}(u_1)\overline{{\cal T}}_{n}(v_1)\overline{{\cal T}}_{n}(u_2){\cal T}_{n}(v_2)\big>.\label{rhoAn3}
\end{equation}
It is to be noted that $Tr(\rho_A^{T_2})^n$  shows different functional dependence on $|\lambda_i|$ ($\lambda_i$'s are the eigenvalues of $\rho_A^{T_2}$) depending on parity of $n$. Therefore, the required expression for the entanglement negativity may be obtained as an analytic continuation of the even sequences $n$ to $n_e\to1$ (where $n_e$ represents even values of $n$) \cite{Calabrese:2014yza}. Thus, by making use of the replica technique given in eq.(\ref{rhoAn3}), the authors defined the entanglement negativity for the bipartite mixed state of two disjoint intervals in a $CFT_{1+1}$ as  
\begin{eqnarray}
{\cal E}&=&\lim_{n_e \to 1}\ln(Tr[(\rho_A^{T_2})^{n_e}])\\&=&\lim_{n_e \to 1}\ln \left[\big<{\cal T}_{n_e}(u_1)\overline{{\cal T}}_{n_e}(v_1)\overline{{\cal T}}_{n_e}(u_2){\cal T}_{n_e}(v_2)\big>\right].~~\label{ENCFTA}
\end{eqnarray}

\subsection{Entanglement negativity for the bipartite  pure vacuum state}
Here we explain the systematic method developed by the authors in \cite{Calabrese:2012ew,Calabrese:2012nk} in order to obtain the entanglement negativity for the bipartite ($A\cup A^c$) pure state described by the  $CFT_{1+1}$ vacuum. In order to reduce a tripartite system $(A_1,A_2,A^c)$ to a bipartite configuration 
$(A,A^c,\cancel{0})$, the authors make the identification $u_2\rightarrow v_1$ and $v_2\rightarrow u_1$ in eq.(\ref{ENCFTA}) such that the interval corresponding to the subsystem $A$ is now a single interval denoted by $[u,v]$. With this identification, the correct form for the entanglement negativity of the subsystem $A$ is given in terms of the two point twist correlator as 
\begin{equation}
 {\cal E}=\lim_{n_e \to 1}\ln\left[Tr(\rho^{T_A})^{n_e}\right]=\lim_{n_e \to 1}\ln \left[\big<{\cal T}^2_{n_e}(u)\overline{{\cal T}}^2_{n_e}(v)\big>\right],\label{EN0T} 
\end{equation}
where, $\rho=\rho_{A\cup A^c}$ corresponds to the density matrix of the full system. In order to compute the two point twist correlator given in the equation above, the authors in \cite{Calabrese:2014yza} use the fact that the operator ${\cal T}^2_{j}$ connects the $j$-th sheet of the Riemann surface to the $(j+2)$-th sheet . When the parity of $n$ is even i.e $n=n_e$, the $n_e$-sheeted Riemann surface dissociates into two $n_e/2$ sheeted Riemann surfaces which simplifies the expression for the entanglement negativity in eq.(\ref{EN0T}) as follows
\begin{equation}
 {\cal E}=\lim_{n_e \to 1}\ln \left[(\big<{\cal T}_{n_e/2}(u)\overline{{\cal T}}_{n_e/2}(v)\big>)^2\right].\label{ENCFTPA}
\end{equation}
Here the scaling dimension-$\Delta_{n_e}^{(2)}$ of the operator ${\cal T}_{n_e}^2$ is related to the scaling dimension-$(\Delta_{n_e})$ of the operator ${\cal T}_{n_e}$ as 
\begin{eqnarray}
\Delta_{n_e}^{(2)}=2 \Delta_{n_e/2}&=&\frac{c}{6}\left(\frac{n_e}{2}-\frac{2}{n_e}\right),\nonumber\\
\Delta_{n_e}&=& \frac{c}{12}\left(n_e-\frac{1}{n_e}\right).\label{dimTn21}
\end{eqnarray}
Since the form of the two point twist correlator in eq.(\ref{ENCFTPA}) is fixed in a $CFT_{1+1}$, it follows that the expression for the entanglement negativity is given as follows
\begin{equation}
{\cal E}=\frac{c}{2}\ln\left(\frac{\ell}{a}\right)+constant,\label{ENCFTzero}
\end{equation}
where, $\ell=\mid u-v\mid$ is the length of the subsystem-A and $a$ is the UV cutoff for the $(1+1)$- dimensional conformal field theory. From the above discussion one may observe that for the pure state described by the $CFT_{1+1}$ vacuum, the entanglement negativity is equal to the R\'{e}nyi entropy of order-$1/2$ which is a well known result in quantum information theory \cite{PhysRevA.65.032314,Calabrese:2012nk}. 

\subsection{ Entanglement negativity for the bipartite finite temperature mixed state }\label{bfms}
In this section, we review the procedure for the computation of  entanglement negativity for the finite temperature mixed state of a $CFT_{1+1}$ as described in \cite{Calabrese:2014yza}. Note that the  method for obtaining the entanglement negativity for the finite temperature mixed state is subtle and the authors in \cite{Calabrese:2014yza} demonstrated that the naive application of eq.(\ref{EN0T}) is incorrect. The reason for this subtlety may be associated with the fact that the decoupling of the $n_e$ sheeted Riemann surface into two $n_e/2$ sheeted Riemann surfaces leads to a simplified expression for the entanglement negativity given by eq.(\ref{ENCFTPA}). The authors showed that this simplification is suitable only for the pure state scenario when the $CFT_{1+1}$ is on the complex plane. For the finite temperature bipartite mixed state  where the partial transpose is over an infinite cylinder, the expression in eq.(\ref{ENCFTPA}) is unsuitable to compute the entanglement negativity. The  authors in \cite{Calabrese:2014yza} noted that the entanglement negativity of the bipartite ($A\cup A^c$) finite temperature mixed state of a $CFT_{1+1}$ is  related to the following four point twist correlator 
\begin{eqnarray}
{\cal E}&=&\lim_{L \to \infty}\lim_{n_e \to 1}\ln\left[Tr(\rho^{T_A})^{n_e}\right]\\&=&\lim_{L \to \infty}\lim_{n_e \to 1}\ln\left[\big<{\cal T}_{n_e}(-L)\overline{{\cal T}}^2_{n_e}(-\ell){\cal T}^2_{n_e}(0)\overline{{\cal T}}_{n_e}(L)\big>_{\beta}\right]\nonumber\\.~\label{ENCFT1}
\end{eqnarray}
In the above equation, the interval corresponding to subsystem-$A$ is given by $[u,v]=[-\ell,0]$ whereas, ${\cal T}_{n_e}(-L)$ and  ${\cal T}_{n}(L)$ correspond to the twist fields located at the end points of the subsystems denoted as $B_1=[-L,-\ell]$ and $B_2=[0,L]$ at some large distance $L$ from the interval $A$. Moreover, if we denote $B=B_1 \cup B_2$ then the the limit $L\to \infty$ in eq.(\ref{ENCFT1}) corresponds to $B\to A^c$. Here, it is also to be noted that in order to get the correct result from eq.(\ref{ENCFT1}), the limit $(L \rightarrow \infty)$ should be applied only after taking the replica limit $(n_e\rightarrow 1)$.  The subscript $\beta$ indicates that at finite temperatures it is required to evaluate the four point function in eq.(\ref{ENCFT1}) on an infinitely long cylinder of circumference $\beta=1/T$. This cylindrical geometry may be obtained from the 2-dimensional complex plane by the following conformal transformation
\begin{equation}
z\rightarrow \omega=\frac{\beta}{2 \pi}\ln z,\label{ct}
\end{equation}
where, $z$ denotes the coordinates on the complex plane and $\omega$ denotes the coordinates on the cylinder.  Under the conformal transformation given by eq.(\ref{ct}), the required four-point function of a $\it{CFT_{1+1}} $ on the infinite cylinder is related to the four point function on the complex plane as follows
 \begin{equation}
 \begin{split}
  &\big<{\cal T}_{n_e}(w_1)\overline{{\cal T}}^2_{n_e}(w_2){\cal T}^2_{n_e}(w_3)\overline{{\cal T}}_{n_e}(w_4)\big>_{\beta}\\&=\prod_j{|z'(w_j)|^{\Delta_j}}  \big<{\cal T}_{n_e}(z_1)\overline{{\cal T}}^2_{n_e}(z_2){\cal T}^2_{n_e}(z_3)\overline{{\cal T}}_{n_e}(z_4)\big>_{\mathbb{C}},\label{t4pt}
 \end{split}
 \end{equation}
here $z'(w_j)=\frac{dz}{dw}|_{z=w_j}$ and $\Delta_j$ is the scaling dimension of operator inserted at $w_j$. The form of the four point twist correlator on the complex plane is given as follows
\begin{equation}
\begin{split}
 &\big<{\cal T}_{n_e}(z_1)\overline{{\cal T}}^2_{n_e}(z_2){\cal T}^2_{n_e}(z_3)\overline{{\cal T}}_{n_e}(z_4)\big>_{\mathbb{C}}\\&=\frac{1}{z_{14}^{2\Delta_{n_e}}z_{23}^{2\Delta^{(2)}_{n_e}}}\frac{{\cal G}_{n_e}(x)}{x^{\Delta_{n_e}+\Delta^{(2)}_{n_e}}},\label{4ptcft}                
\end{split}
\end{equation}
where the cross ratio $x=\frac{z_{12}z_{34}}{z_{13}z_{24}}$.
In the above equation the $z_i$'s correspond to arbitrary complex numbers such that $z_{ij}=|z_i-z_j|$ with $\big<.\big>$ standing for the expectation value. From eq.(\ref{4ptcft}) it may be observed that the four point twist correlator is only fixed up to an undetermined function ${\cal G}_{n_e}(x)$ of the cross-ratio $x$. The cross ratio $x$ of the four points has two limits $x\rightarrow 0$ and $x\rightarrow 1$, which correspond to high and low temperature limits respectively \cite{Calabrese:2014yza}. The behavior of the four point function mentioned above at low and high temperatures may be obtained through the OPE of ${\cal T}_{n_e}(u)\overline{{\cal T}}_{n_e}(v)$, ${\cal T}_{n_e}^2(u)\overline{{\cal T}}_{n_e}^2(v)$ and ${\cal T}_{n_e}(u)\overline{{\cal T}}_{n_e}^2(v)$. For low temperatures one has $x\to1$ i.e $z_3\rightarrow z_2$, $z_4\rightarrow z_1$ which leads to the following form of the four point correlator in eq.(\ref{4ptcft})
 \begin{equation}
  \big<{\cal T}_{n_e}(z_1)\overline{{\cal T}}^2_{n_e}(z_2){\cal T}^2_{n_e}(z_3)\overline{{\cal T}}_{n_e}(z_4)\big>= \big<{\cal T}_{n_e}(z_1)\overline{{\cal T}}_{n_e}(z_4)\big>\big<{\cal T}^2_{n_e}(z_2)\overline{{\cal T}}^2_{n_e}(z_3)\big>+...\ .\label{4ptlt}
 \end{equation}
On the other hand the high temperatures limit is given by $x\rightarrow 0$ i.e $z_2\rightarrow z_1$, $z_4\rightarrow z_3$, which  results in the following form for the four point twist correlator
\begin{equation}
 \big<{\cal T}_{n_e}(z_1)\overline{{\cal T}}^2_{n_e}(z_2){\cal T}^2_{n_e}(z_3)\overline{{\cal T}}_{n_e}(z_4)\big>=\frac{C_{n_e}^2 c_{n_e}}{(z_{12}z_{34})^{\Delta^{(2)}_{n_e}}z_{13}^{2\Delta_{n_e}}}+...\label{4ptht}
\end{equation}
Here $c_{n_e}$ and $C_{n_e}$ are constants that appear as the coefficients of the leading term in the OPE of the two point twist correlators ${\cal T}_{n_e}(u)\overline{{\cal T}}_{n_e}(v)$ and  ${\cal T}_{n_e}(u)\overline{{\cal T}}_{n_e}^2(v)$ respectively. The high and low temperature behavior  given in eq.(\ref{4ptlt}) and eq.(\ref{4ptht}) leads to following suggestive form for the four point correlator 
\begin{equation}
 \big<{\cal T}_{n_e}(z_1)\overline{{\cal T}}^2_{n_e}(z_2){\cal T}^2_{n_e}(z_3)\overline{{\cal T}}_{n_e}(z_4)\big>_{\mathbb{C}}=\frac{c_{n_e}c^2_{n_e/2}}{z_{14}^{2\Delta_{n_e}}z_{23}^{2\Delta^{(2)}_{n_e}}}\frac{{\cal F}_{n_e}(x)}{x^{\Delta^{(2)}_{n_e}}},\label{rhoAn5}
\end{equation}
where, $c_{n_e}$ and $c^2_{n_e/2}$ are constants. Following \cite{Calabrese:2014yza} , one may also obtain the constraints on the function ${\cal F}_{n_e}(x)$ in the two limits $x\to 1$ and $x \to 0$  as follows
\begin{equation}
{\cal F}_{n_e}(1)=1,~~~{\cal F}_{n_e}(0)=\frac{C_{n_e}^2}{c^2_{n_e/2}}.\label{cons1}
\end{equation}
Rewriting $z_i$'s in eq.(\ref{rhoAn5}) in terms of the required coordinates on the infinite cylinder i.e $(z_1,z_2,z_3,z_4)\rightarrow(e^{-2\pi L/\beta},e^{-2\pi \ell/\beta},1,e^{2\pi L/\beta})$  and then using the transformation given by eq.(\ref{t4pt}) one may obtain the required four point correlator. Thus, the entanglement negativity for the bipartite ($A\cup A^c$) finite temperature mixed state of a  $CFT_{1+1}$ due to Calabrese et al. \cite{Calabrese:2014yza} may be expressed as follows
\begin{equation}
 {\cal E}=\frac{c}{2}\ln\left[\frac{\beta}{\pi a}\sinh\left(\frac{\pi \ell}{\beta}\right)\right]-\frac{\pi c \ell}{2\beta}+f(e^{-2\pi \ell/\beta})+\ln (c^2_{1/2} c_1).\label{ENCFTfinite}
\end{equation}
The function $f(x)$ in the above expression is defined in the replica limit $(n_e\rightarrow 1)$ as follows
\begin{equation}
f(x)=\lim_{n_e \to 1}\ln[{\cal F}_{n_e}(x)],~~~~~\lim_{L \to \infty}x=e^{-2\pi \ell/\beta}
\end{equation}
Note that the second term in the eq.(\ref{ENCFTfinite}) corresponds to the thermal entropy of the subsystem $A$ up to a numerical factor. Therefore, eq.(\ref{ENCFTfinite}) clearly indicates that the entanglement negativity characterizes the distillable entanglement for the finite temperature mixed state of a $CFT_{1+1}$ through the elimination of the contribution from the thermal correlations.

\end{appendices}

\bibliographystyle{utphys} 
\bibliography{H_ENbib1} 

\end{document}